\documentclass[12pt, a4paper]{article}

\usepackage[latin1]{inputenc}
\usepackage[dvips]{graphicx,color}
\usepackage{amsmath}
\usepackage{amsfonts}
\usepackage{amssymb}
\usepackage{rotating}
\numberwithin{equation}{section}

\begin{document}

\def\pplogo{\vbox{\kern-\headheight\kern -29pt
\halign{##&##\hfil\cr&{\ppnumber}\cr\rule{0pt}{2.5ex}&\ppdate\cr}}}
\makeatletter
\def\ps@firstpage{\ps@empty \def\@oddhead{\hss\pplogo}%
  \let\@evenhead\@oddhead 
}

\def\maketitle{\par
 \begingroup
 \def\thefootnote{\fnsymbol{footnote}}
 \def\@makefnmark{\hbox{$^{\@thefnmark}$\hss}}
 \if@twocolumn
 \twocolumn[\@maketitle]
 \else \newpage
 \global\@topnum\z@ \@maketitle \fi\thispagestyle{firstpage}\@thanks
 \endgroup
 \setcounter{footnote}{0}
 \let\maketitle\relax
 \let\@maketitle\relax
 \gdef\@thanks{}\gdef\@author{}\gdef\@title{}\let\thanks\relax}
\makeatother

\def\eq#1{Eq. (\ref{eq:#1})}
\newcommand{\nc}{\newcommand}
\def\theequation{\thesection.\arabic{equation}}
\nc{\beq}{\begin{equation}}
\nc{\eeq}{\end{equation}}
\nc{\barray}{\begin{eqnarray}}
\nc{\earray}{\end{eqnarray}}
\nc{\barrayn}{\begin{eqnarray*}}
\nc{\earrayn}{\end{eqnarray*}}
\nc{\bcenter}{\begin{center}}
\nc{\ecenter}{\end{center}}
\nc{\ket}[1]{| #1 \rangle}
\nc{\bra}[1]{\langle #1 |}
\nc{\0}{\ket{0}}
\nc{\mc}{\mathcal}
\nc{\etal}{{\em et al}}
\nc{\GeV}{\mbox{GeV}}
\nc{\er}[1]{(\ref{eq:#1})}
\nc{\onehalf}{\frac{1}{2}}
\nc{\partialbar}{\bar{\partial}}
\nc{\psit}{\widetilde{\psi}}
\nc{\Tr}{\mbox{Tr}}
\nc{\tc}{\tilde c}
\nc{\tk}{\tilde K}
\nc{\tv}{\tilde V}
\nc{\CN}{{\mathcal N}}
\nc{\muphi}{\mu_{\phi}}

\def\beg{\begin{equation}}
\def\eq{\end{equation}}


\setcounter{page}0
\def\ppnumber{\vbox{\baselineskip14pt
\hbox{\footnotesize{RUNHETC-2009-16}}
}}
\def\ppdate{}\date{}
\author{Jean-Fran\c{c}ois Fortin, Jessie Shelton, Scott Thomas and Yue Zhao\\
[7mm]
{\normalsize New High Energy Theory Center}\\
{\normalsize Department of Physics and Astronomy}\\
{\normalsize Rutgers University, Piscataway, NJ 08854--8019, USA}\\
}

\title{\bf \vskip -1.2cm Gamma Ray Spectra from\\
[1mm] Dark Matter Annihilation and Decay \vskip 0.2cm} \maketitle

\begin{abstract} \small
\noindent

In this paper, we study gamma ray spectra for various scenarios of
dark matter annihilation and decay.  We focus on processes which
generate only high-energy photons or leptons and photons, but no
proton-antiproton pairs, to be compatible with PAMELA's data.  We
investigate photons produced directly from two-body decay chains and
photons produced together with charged particles.  For the former
case we also include the process $\mbox{DM}(+\mbox{DM})\rightarrow
N\phi\rightarrow2N\gamma$ which can arise from specific
strongly-coupled dark matter scenarios.  For the latter case,
photons are either generated by final state radiation from
high-energy leptons or are directly generated from contact interactions
represented by higher-order (non-renormalizable) operators obtained after
integrating out heavy modes.  We compare their overall annihilation cross-sections/decay
rates taking into account chiral suppression (in the s-wave
approximation), dimension of operators and dark matter particle
properties.  A rough estimate shows that, for a dark matter particle
with a mass of $\mathcal{O}(1\,{\rm TeV})$, the hard photon spectra in direct
electron-positron-photon final states arising from either
scalar boson dark matter annihilation/decay or Majorana fermion dark
matter annihilation
are dominated by higher-order operators if the
scale of the leading operator is lower than $\mathcal{O}(1000\,{\rm
TeV})$.  Otherwise, all the photon spectra arising in this way
are dominated by final state radiation.  Among the spectra
studied, the higher-order operators spectrum is the hardest while
the final state radiation spectrum with an intermediate decay
is the softest.

\end{abstract}
\bigskip
\newpage

\tableofcontents

\vskip 1cm


\newpage

\section{Introduction}\label{sec:intro}


Recently several experiments (ATIC \cite{:2008zzr}, H.E.S.S.
\cite{Collaboration:2008aaa,Aharonian:2009ah}, PAMELA
\cite{Adriani:2008zr}, FERMI \cite{Abdo:2009zk}) detected an anomaly
in the cosmic ray electron spectrum, measuring an excess in the
high-energy positron flux compare to usual diffusion models.  Such
an excess was not found for antiprotons \cite{Adriani:2008zq}.  One
was thus led to conjecture a new source of primary electrons and
positrons.  The most likely sources advanced to explain this anomaly
are nearby pulsars or dark matter (DM) annihilation/decay
\cite{Cirelli:2008pk,ArkaniHamed:2008qn}.  In any case, the issue
related to the identity of the new source (pulsars versus DM) will
be clarified by the photon spectrum which will be measured by FERMI
and announced later this summer.  Indeed, there is an irreducible
background of gamma ray photons coming from final state radiation
(FSR) from electrons and/or positrons.  Nearby pulsars would lead to
a local photon spectrum while DM would instead generate a diffuse
photon spectrum.

The existence of DM is well established, although its paticle identity is
still unknown.  Assuming that DM annihilation/decay is the new
source of primary electrons and positrons, several scenarios are
possible
(see [10--33] for related work).
Apart from the FSR of photons, DM might also
annihilate/decay directly to photons.  Different photon spectra are
expected from each annihilation/decay mode and a knowledge of the
achievable spectra will help physicists understand the properties of
DM at the particle level.

In Section \ref{sec:direct}, we discuss direct production of photons
from DM annihilation/decay through subsequent two-body decay chain.
In Section \ref{sec:irr}, we study the irreducible photon background
from charged particles, considering only the dominant process
(either FSR of photons or photon production from higher-order
operators).  Next, in Section \ref{sec:taus} we consider direct
photon production from taus.  Finally, Section \ref{sec:photon}
contains a summary of the results and a comparison of the different
photon spectra and total fluxes obtained from the density of states
(DOS).  Various detailed computations are left for the Appendix.

Since we want to address both annihilation and decay, we compute
photon DOS throughout the paper, and it only differs with the
spectrum by a constant factor as total number of photons.  One can
write DOS as
\begin{equation*}
\frac{1}{N_{\gamma}}\frac{dN_\gamma}{dE_\gamma}=\frac{1}{\langle\sigma
v\rangle}\frac{d\langle\sigma
v\rangle}{dE_\gamma}\hspace{0.5cm}\mbox{or}\hspace{0.5cm}\frac{1}{N_{\gamma}}\frac{dN_\gamma}{dE_\gamma}=\frac{1}{\Gamma}\frac{d\Gamma}{dE_\gamma},
\end{equation*}
\textit{not} photon spectra $\frac{dN_\gamma}{dE_\gamma}$.  The
photon DOS are thus normalized to 1, i.e. $\int
dE_\gamma\frac{1}{N_{\gamma}}\frac{dN_\gamma}{dE_\gamma}=1$, except
for FSR where the DOS is normalized with respect to the annihilation
cross-section/decay to zeroth-order in the fine structure constant
(see Section \ref{sec:irr}). Thus, the total number of photons
$N_\gamma$, given by $N_\gamma=\int
dE_\gamma\frac{dN_\gamma}{dE_\gamma}$, can be found from the photon
multiplicity.  For example, a scenario, where DM annihilates/decays
to two scalar bosons which subsequently decay to two photons each,
would have a total number of photons given by $N_\gamma=4$.

Since we focus on the photon DOS, all the results are applicable to
DM annihilation and decay, and both are assessed simultaneously by
substituting the parameter $M$ by $2m_{\rm DM}$ for DM annihilation
or $m_{\rm DM}$ for DM decay.  Since backgrounds fall roughly like
$E_\gamma^2$, we plot the photon DOS in function of the
dimensionless photon energy $2E_\gamma/M$ with an extra $E_\gamma^2$
factor.  This is consistent with the standard representation used in
experiments.  The $M/2$ factor appearing in the plots is to make the
quantities dimensionless, where $M/2$ is the maximal energy a photon
can get during a process.

Throughout this paper we assume that DM annihilates/decays only to
leptons or photons as indicated by the experiments. Moreover, we
assume that DM annihilation always occurs in the s-wave
approximation.  Under this assumption, we can get rid of the
operators that can only contribute in p wave or higher order waves
when we do operator analysis in later sections. Furthermore, DM
annihilation to leptons or photons is allowed whatever the particle
identity of DM (scalar boson, fermion or gauge boson). However, with
the assumptions that standard model particles are not charged under
hidden symmetries and that individual lepton numbers are conserved,
DM decay to leptons or photons is allowed only for scalar boson and
abelian gauge boson DM.

Issues related to the overall annihilation cross-sections/decay
rates (Sommerfeld enhancement, non-thermal DM production, etc) and
the irreducible astrophysical photon background (inverse Compton
scattering from starlight and CMB, synchrotron radiation from
galactic magnetic fields) will not be investigated.


\section{Direct production of photons through subsequent two-body decay chain}\label{sec:direct}

In this section, we analyze direct production of photons through
subsequent two-body decay chain produced by DM annihilation/decay.
We assume the whole process is a chain with $k$ steps,
$\phi_{i-1}\rightarrow2\phi_i$, the last $\phi$ will decay to two
photons and cause $2k$ photons as the final products.  In Appendix
\ref{app:DOS}, we give a general way to calculate the DOS for
two-body decay chains with on-shell intermediate particles.  In
following subsections, we simply show some results for different
scenarios.

\subsection{$\mbox{DM}+\mbox{DM}\rightarrow2\gamma$ and $\mbox{DM}\rightarrow2\gamma$}\label{subsec:2photons}

We first start with the simplest case where DM annihilates/decays
directly to two photons.  Since DM particles are almost stationary
in the galactic frame, the photon DOS is a pure delta function,
\begin{equation}\label{eqn:2photons}
\frac{1}{N_{\gamma}}\frac{dN_\gamma}{dE_\gamma}=\delta\left(E_\gamma-\frac{M}{2}\right)
\end{equation}
Since $N_\gamma=2$ the photon spectrum is simply twice the photon
DOS.

\subsection{$\mbox{DM}+\mbox{DM}\rightarrow2\phi$ and $\mbox{DM}\rightarrow2\phi$ followed by $\phi\rightarrow2\gamma$}\label{subsec:4photons}

To compute the photon DOS when DM annihilates/decays to two bosons which subsequently decay to two photons each, we must first boost the DOS obtained in the previous subsection and convolute it with the appropriate DOS of DM annihilation/decay to two bosons as explained in Appendix \ref{app:DOS}.  Because the energy of the $\phi$ boson is always $\frac{M}{2}$ and the direction of the photon in the boson rest frame is uniform, we can simply boost the delta function into the DM center of mass frame to obtain the photon DOS,
\begin{equation}\label{eqn:4photons}
\frac{1}{N_{\gamma}}\frac{dN_\gamma}{dE_\gamma}=\frac{2}{M\sqrt{1-\frac{4m_\phi^2}{M^2}}}
\end{equation}
where the photon energy is between
\begin{equation*}
\frac{M}{4}\left(1-\sqrt{1-\frac{4m_\phi^2}{M^2}}\right)<E_\gamma<\frac{M}{4}\left(1+\sqrt{1-\frac{4m_\phi^2}{M^2}}\right)
\end{equation*}
and $m_\phi$ is the boson mass.  As expected the photon DOS is
normalized to one and the photon spectrum is four times the photon
DOS ($N_\gamma=4$).  Notice that in the limit where
$m_\phi=\frac{M}{2}$, in which case the two $\phi$ bosons are
produced \textit{at rest} in the DM center of mass frame, the photon
DOS becomes a delta function and matches the photon DOS obtained in
the previous subsection.  This is easily understood since in this
limit the $\phi$ bosons are stationary and decay to two photons each
as in the previous subsection, only the delta function support and
the DOS normalization change.

\subsection{$\mbox{DM}+\mbox{DM}\rightarrow2\phi$ and $\mbox{DM}\rightarrow2\phi$ followed by $\phi\rightarrow2\pi$ and $\pi\rightarrow2\gamma$}\label{subsec:8photons}

\begin{figure}[ht]
\begin{center}
\includegraphics[scale=1.2]{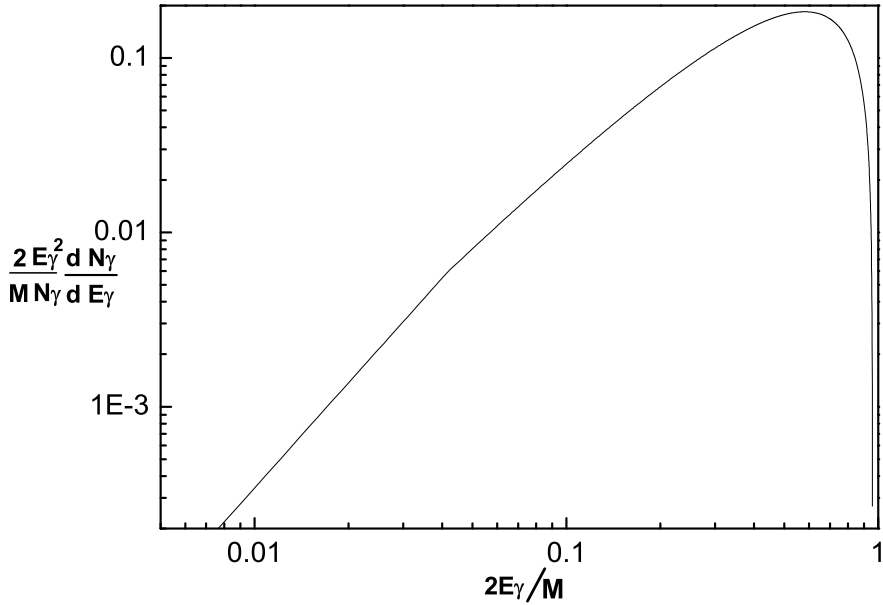}
\caption{Photon spectral distribution for $\mbox{DM}+\mbox{DM}\rightarrow2\phi$ and
$\mbox{DM}\rightarrow2\phi$ followed by $\phi\rightarrow2\pi$ and
$\pi\rightarrow2\gamma$ with $M=2000$ GeV, $m_\phi=400$ GeV and
$m_\pi=0.14$ GeV.
The distributions peaks at
$2E_\gamma/M= E_{\gamma}/E_{\gamma}^{\rm max}=
(E_\pi^{\rm max}+|\vec{p}_\pi^{\;\rm max}|)/Me^{1/2}$,
where $e \simeq 2.718$; for the parameters here the peak is at
$2E_\gamma/M \simeq 0.581$. } \label{fig:Tripledecaychain}
\end{center}
\end{figure}

When DM annihilates/decays to two $\phi$ bosons which decay to two $\pi$ bosons and finally decay to two photons, the photon DOS can again be computed by boosting the DOS obtained in the previous subsection and convoluting the result with the appropriate DOS of DM annihilation/decay to two bosons.  As shown before, the $\phi$ DOS is a delta function and the $\pi$ DOS is a step function.  However the photon DOS gets complicated since, for a fixed photon energy, not all $\pi$ with energy in the support of the step function can contribute.  Indeed we have to calculate the minimum and maximum allowed $\pi$ energies which can generate the relevant photon energy and compute the photon DOS accordingly.  Detailed computations are discussed in Appendix \ref{app:DOS}.  Defining the boson masses as $m_\phi$ and $m_\pi$ the photon DOS is
\begin{equation}\label{eqn:8photons}
\frac{1}{N_{\gamma}}\frac{dN_\gamma}{dE_\gamma}=A\left\{
\begin{array}{ll}
\ln\left[\frac{E_\pi^{\rm max}+|\vec{p}_\pi^{\;\rm max}|}{m_\pi}\frac{2E_\gamma}{m_\pi}\right] & \mbox{for $E_\gamma^{\rm min}(E_\pi^{\rm max})<E_\gamma<E_\gamma^{\rm min}(E_\pi^{\rm min})$}\\
\ln\left[\frac{E_\pi^{\rm max}+|\vec{p}_\pi^{\;\rm max}|}{E_\pi^{\rm min}+|\vec{p}_\pi^{\;\rm min}|}\right] & \mbox{for $E_\gamma^{\rm min}(E_\pi^{\rm min})<E_\gamma<E_\gamma^{\rm max}(E_\pi^{\rm min})$}\\
\ln\left[\frac{E_\pi^{\rm max}+|\vec{p}_\pi^{\;\rm max}|}{m_\pi}\frac{m_\pi}{2E_\gamma}\right] & \mbox{for $E_\gamma^{\rm max}(E_\pi^{\rm min})<E_\gamma<E_\gamma^{\rm max}(E_\pi^{\rm max})$}
\end{array}\right.
\end{equation}
where
\begin{eqnarray*}
A &=& \frac{2}{M\sqrt{1-\frac{4m_\phi^2}{M^2}}\sqrt{1-\frac{4m_\pi^2}{m_\phi^2}}}\\
E_\pi^{\rm min} &=& \frac{M}{4}\left(1-\sqrt{1-\frac{4m_\phi^2}{M^2}}\sqrt{1-\frac{4m_\pi^2}{m_\phi^2}}\right)\\
E_\pi^{\rm max} &=& \frac{M}{4}\left(1+\sqrt{1-\frac{4m_\phi^2}{M^2}}\sqrt{1-\frac{4m_\pi^2}{m_\phi^2}}\right)
\end{eqnarray*}
and $|\vec{p}_\pi|=\sqrt{E_\pi^2-m_\pi^2}$.  The limits on the photon energy $E_\gamma$ are found using
\begin{eqnarray*}
E_\gamma^{\rm min}(E_\pi) &=& \frac{E_\pi}{2}\left(1-\sqrt{1-\frac{m_\pi^2}{E_\pi^2}}\right)\\
E_\gamma^{\rm max}(E_\pi) &=& \frac{E_\pi}{2}\left(1+\sqrt{1-\frac{m_\pi^2}{E_\pi^2}}\right)
\end{eqnarray*}
where $E_\gamma^{\rm min}(E_\pi^{\rm max})<E_\gamma^{\rm
min}(E_\pi^{\rm min})<E_\gamma^{\rm max}(E_\pi^{\rm
min})<E_\gamma^{\rm max}(E_\pi^{\rm max})$.  $E_\gamma^{\rm
min/max}(E_\pi^{\rm min})$ is the minimum/maximum photon energy that
can be generated by a $\pi$ boson with minimum energy and similarly
for $E_\gamma^{\rm min/max}(E_\pi^{\rm max})$.  Again the photon DOS
is canonically normalized while the photon spectrum is normalized
such that the total number of photons is 8 ($N_\gamma=8$) and it is
shown in figure \ref{fig:Tripledecaychain} for some given boson
masses.

There are three parts in DOS.  Firstly, DOS increases with photon
energy as a log function, then in the second part DOS is just a
constant, does not depend on $E_\gamma$, and in the third part DOS
will decrease with $E_\gamma$ logarithmically.  Since the parameters
we take for the figure, the first part is out of region, the second
part is corresponding to a straight line at low energy regime, and
the joint point of second and third part will cause a kink at
$2E_\gamma/M\approx0.04$ since they are not smoothly connected to
each other.

As a consistency check, we can again match the DOS obtained here
with the DOS obtained in the previous subsections by taking
different limits on the mass ratios.  Indeed, by taking the limits
$\frac{m_\pi}{m_\phi}=\frac{1}{2}$ or $\frac{m_\phi}{M}=\frac{1}{2}$
(but not both) the DOS becomes a step function because, in such
limits, there are two particles produced \textit{at rest} in the
center of mass frame of the parent particle(s).  Moreover in the
limit where both $\frac{m_\pi}{m_\phi}=\frac{1}{2}$ and
$\frac{m_\phi}{M}=\frac{1}{2}$ the DOS becomes a delta function
since, in that limit, DM annihilates/decays to two stationary $\phi$
bosons, and both $\phi$ decay to two stationary $\pi$ bosons, which
finally decay to a total of eight photons, each of them taking
$\frac{1}{8}$ of the initial energy $M$.

If we take some generic values for the two mass ratios
$\frac{m_\pi}{m_\phi}$ and $\frac{m_\phi}{M}$, the third part of the
DOS, which decreases as $-\ln(E_\gamma)$, dominates the DOS (it has
the largest support).  Also, adding more steps to the decay chain
results in a softer photon DOS, a general trend which is intuitively
expected since the total energy is distributed among a larger number
of particles.

\subsection{$\mbox{DM}+\mbox{DM}\rightarrow N\phi$ and $\mbox{DM}\rightarrow N\phi$ followed by $\phi\rightarrow2\gamma$}\label{subsec:2Nphotons}

\begin{figure}[ht]
\begin{center}
\includegraphics[scale=1.2]{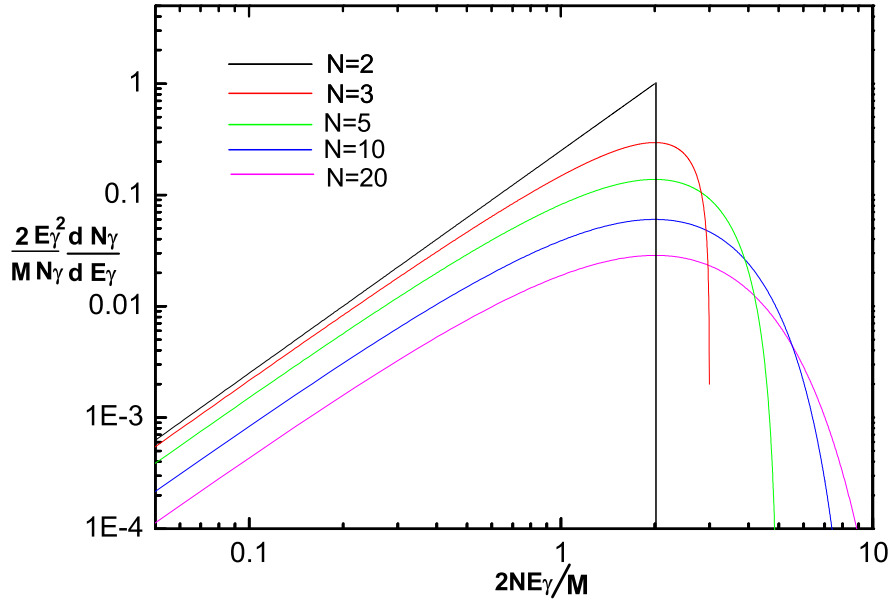}
\caption{Photon spectral distributions for $\mbox{DM}+\mbox{DM}\rightarrow N\phi$ and
$\mbox{DM}\rightarrow N\phi$ followed by $\phi\rightarrow2\gamma$
in the limit $M \gg m_\phi$, for $N=2,3,5,10$ and $20$.
All the distributions peak at $2NE_\gamma/M= E_{\gamma}/(E_{\gamma}^{\rm max}/N) = 2$.}
\label{fig:Nintermediatestates}
\end{center}
\end{figure}

Finally, we study DM annihilation/decay to $N$ $\phi$ bosons which
then decay to two photons each.  The computation is done in the
massless limit, i.e. $m_\phi=0$.  This scenario occurs for example
when two strongly-coupled bound states annihilate to very light
pseudo Nambu-Goldstone bosons (pions in the QCD analogy) which then
decay to two photons \cite{Banks:2009rb}.  In the massless limit,
the $\phi$ boson DOS, assuming constant matrix element, can be
computed by dimensional analysis and is given by
\begin{equation}
\frac{dN_\phi}{dE_\phi}=\frac{2(N-1)(N-2)(M-2E_\phi)^{N-3}2E_\phi}{M^{N-1}}
\end{equation}
where $0<E_\phi<\frac{M}{2}$.  Therefore the photon DOS is simply
\begin{equation}\label{eqn:2Nphotons}
\frac{1}{N_{\gamma}}\frac{dN_\gamma}{dE_\gamma}=\frac{2(N-1)(M-2E_\gamma)^{N-2}}{M^{N-1}}
\end{equation}
where $0<E_\gamma<\frac{M}{2}$ (more detail is given in Appendix
\ref{app:DOS}).  The photon DOS satisfies $n_\gamma=1$ and the
photon spectrum is normalized to $N_\gamma=2N$.  When $N=2$ this
result is equivalent to the scenario discussed in subsection
\ref{subsec:4photons}, equation (\ref{eqn:4photons}), with
$m_\phi=0$.  The photon DOS for different $N$ is shown in figure
\ref{fig:Nintermediatestates}.  One can easily see that the width
grows with N.


\section{Photons from final states with charged particles}
\label{sec:irr}

In this section, we analyze the irreducible photon background coming
from charged particles, considering only the dominant process, i.e.
FSR of photons or direct photon production from higher-order
operators.

For DM annihilation, FSR is the dominant process unless DM is a
scalar boson or a Majorana fermion which annihilates directly to an
electron-positron pair.  For DM decay, FSR is the dominant process
unless DM is a scalar boson which decays directly to an
electron-positron pair.  Indeed, in these specific cases FSR is
small due to the large chiral suppression (in the s-wave
approximation for the annihilation scenarios), and might or might
not be the dominant process according to the typical scale of the
leading higher-order operators.

Here we will study two different modes which contribute to the
photon DOS: DM annihilation/decay to one electron-positron pair and
DM annihilation/decay to one boson pair which subsequently decay to
one electron-positron pair each.

\subsection{$\mbox{DM}+\mbox{DM}\rightarrow e^++e^-+\gamma$ and $\mbox{DM}\rightarrow e^++e^-+\gamma$}\label{subsec:1ee1photon}

\subsubsection*{Final state radiation of photons}

\begin{figure}[ht]
\begin{center}
\includegraphics[scale=1.2]{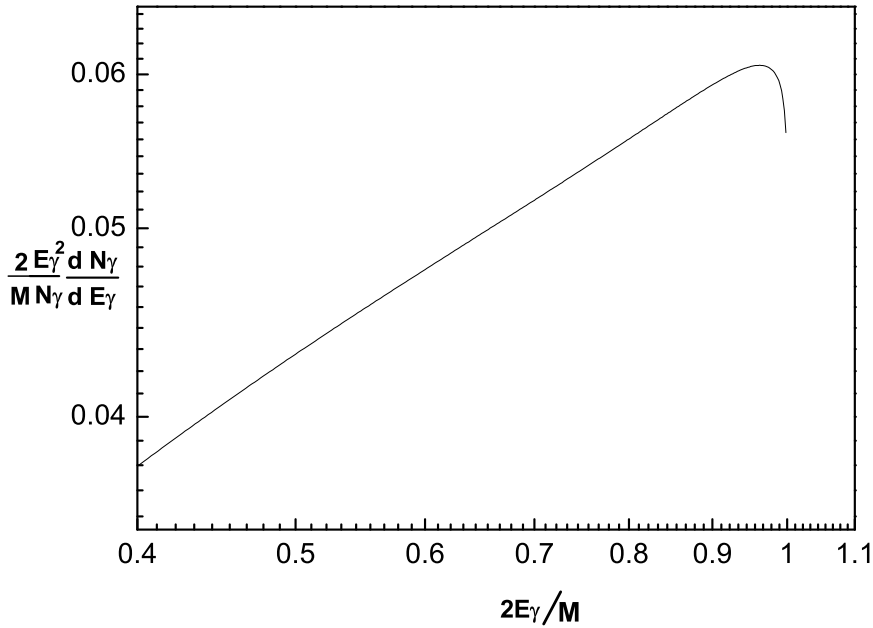}
\caption{Photon spectral distribution for $\mbox{DM}+\mbox{DM}\rightarrow
e^++e^-+\gamma$ and $\mbox{DM}\rightarrow e^++e^-+\gamma$ from FSR
with $M=2000$ GeV.
The distribution peaks at $2E_\gamma/M=E_{\gamma}/E_{\gamma}^{\rm max}= x$
where $x$ is the solution of
${x(x^2-2x+2)}/({(1-x)(3x^2-4x+2)})=\ln\left[{M^2(1-x)}/{m_e^2}\right]$;
for the parameters here the peak is at $2E_\gamma/M  \simeq 0.962$.}
\label{fig:FSR}
\end{center}
\end{figure}

For direct DM annihilation/decay to one electron-positron pair, the
photon DOS from FSR of a single photon from the electron or the
positron, in the collinear limit (a formula at leading order in the
electron mass is given in Appendix \ref{app:FSR}), is simply given
by \cite{Bergstrom:1989jr}
\begin{equation}\label{eqn:FSR}
\frac{1}{N_{\gamma}}\frac{dN_\gamma}{dE_\gamma}\simeq\frac{\alpha}{\pi}\left(\frac{M^2+(M-2E_\gamma)^2}{M^2E_\gamma}\ln\left[\frac{M(M-2E_\gamma)}{m_e^2}\right]\right)
\end{equation}
and the DOS is shown in figure \ref{fig:FSR}.  Since there is soft
photon Log divergence in the FSR, we choose to normalize respect to
zero$th$ order approximation of $\alpha$, i.e.
$\sigma_{DM+DM\rightarrow e^+ + e^-}$ for annihilation and
$\Gamma_{DM\rightarrow e^+ + e^-}$ for decay.
Notice here that the photon spectrum is the same than the photon DOS.

\subsubsection*{Direct photon production from higher-order operators}

Photons from FSR may not always be the leading contribution to the
spectrum in some cases.  The smallness of the electron mass might
lead to large chiral suppression for FSR.  So the leading process
can be the one where photons are directly generated associated with
$e^+ e^-$. In low energy effective theory, one can write such
process as higher order operators, assuming the particle
intermediates the process to be heavy.  We will show below that
there is a low $M_{\rm int}$ regime where the higher-order operators
dominate over FSR for the following scenarios: direct scalar boson
or Majorana DM s-wave annihilation to one electron-positron pair or
direct scalar boson DM decay to one electron-positron pair.

For all these scenarios, named as scalar boson DM annihilation,
Majorana DM annihilation and scalar boson DM decay, in the higher
order operators, they share the common part as
$e^{\dag}\bar{\sigma}^{\mu}e F^{\alpha
\beta}$ or $\bar{e}^{\dag}\sigma^{\mu}\bar{e} F^{\alpha \beta}$. The
only difference among them comes from the part of DM operators. But
that part, under s-wave approximation, will only contribute
different constant coefficients to the spectra of those scenarios.
Since we are calculating DOS instead of spectrum, that difference
will be exactly canceled by normalization.  Thus, they share the
same expression on DOS.

\begin{figure}[ht]
\begin{center}
\includegraphics[scale=1.2]{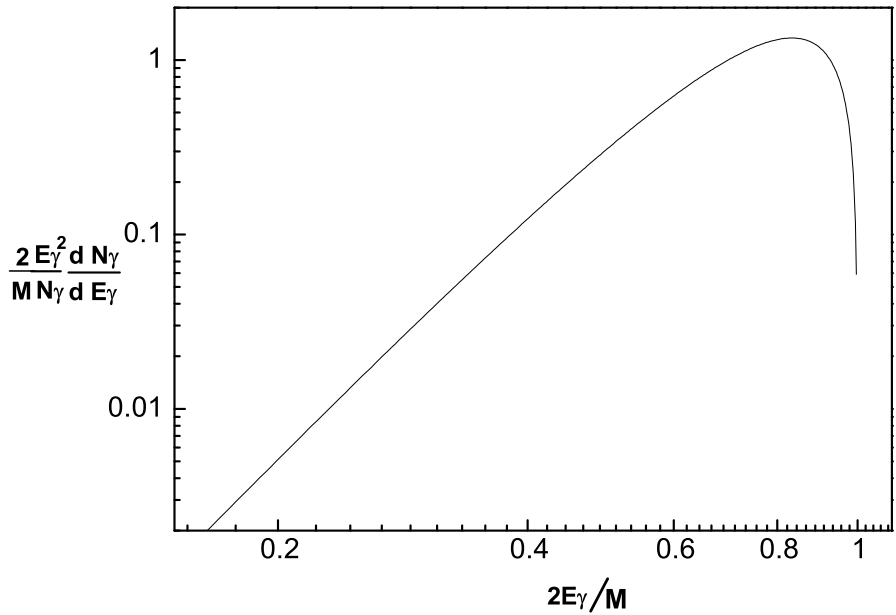}
\caption{Photon spectral distribution for $\mbox{DM}+\mbox{DM}\rightarrow e^++e^-+\gamma$ and
$\mbox{DM}\rightarrow e^++e^-+\gamma$ from the leading
short range contact interaction
for scalar boson or
Majorana fermion $S$-wave annihilation, or scalar boson decay,
in the limit $M \gg m_e$.
The distribution peaks at $2E_\gamma/M=E_{\gamma}/E_{\gamma}^{\rm max} =5/6$.}
\label{fig:Operator}
\end{center}
\end{figure}

Within our assumptions, the
photon DOS from the leading contact interactions given below
in the limit of vanishing electron mass are all given by
\begin{equation}\label{eqn:OPS}
\frac{1}{N_{\gamma}}\frac{dN_\gamma}{dE_\gamma}=\frac{320(M-2E_\gamma)E_\gamma^3}{M^5}
\end{equation}
where $E_{\gamma}^{\rm max} = M/2$.
The photon distribution for this spectrum is shown in figure \ref{fig:Operator}.
The spectrum is independent of
the specific case considered (scalar boson DM annihilation, Majorana
DM annihilation or scalar boson DM decay).  Here the limits on the
photon energy are $0<E_\gamma<\frac{M}{2}$ and DOS is normalized to
1 as well.
Once again, the photon spectrum is the same than the photon DOS in this case.

\subsubsection*{Final state radiation versus direct photon production}

When scalar boson DM $\phi$ annihilates directly to one electron-positron pair,
higher-order operators dominate for small $M_{\rm int}$ while FSR dominates for large $M_{\rm int}$.
Indeed, FSR is generated mainly by the dimension 6 operator
\begin{equation}
\mathcal{L}_{\rm FSR}\supset\frac{hm_e}{M_{\rm int}^2}\phi^\dagger\phi(\bar{e}e+\bar{e}^\dagger e^\dagger)
\end{equation}
while direct photon production is generated from dimension 8 operators through the following effective Lagrangian
\begin{equation}
\mathcal{L}_{\rm eff}\supset\frac{\sqrt{4\pi\alpha}}{M_{\rm int}^2}\partial^\mu(\phi^\dagger\phi)(e^\dagger\bar{\sigma}^\nu e)[a_LF_{\mu\nu}+b_L\tilde{F}_{\mu\nu}]+\{L\rightarrow R,e\rightarrow\bar{e}\}
\end{equation}
where two-component spinor notation \cite{Dreiner:2008tw} is used and the coupling
constants $\{h,a_L,b_L,a_R,b_R\}$ should naturally be order one numbers.
A similar operator has also been considered in Ref \cite{barger}.
The relevant cross-sections are $\langle\sigma_{\rm FSR}v\rangle\approx c_{\rm FSR}\alpha\frac{m_e^2}{M_{\rm int}^4}\ln\left(\frac{4m_\phi^2}{m_e^2}\right)$ and $\langle\sigma_{\rm eff}\rangle\approx c_{\rm eff}\alpha\frac{(2m_\phi)^6}{M_{\rm int}^8}$ where $c_{\rm FSR}\propto h^2$ and $c_{\rm eff}\propto a_L^2+\cdots$ include the appropriate order one coupling constants together with the $\pi$ factors from the phase space integration.

When Majorana DM $\chi$ annihilates directly to one electron-positron pair, higher-order operators dominate for small $M_{\rm int}$ while FSR dominates for large $M_{\rm int}$.  Indeed, FSR is generated mainly by the dimension 6 operator
\begin{equation}
\mathcal{L}_{\rm FSR}\supset\frac{h_L}{M_{\rm int}^2}(\chi^\dagger\bar{\sigma}^\mu\chi)(e^\dagger\bar{\sigma}_\mu e)+\{L\rightarrow R,e\rightarrow\bar{e}\}
\end{equation}
while direct photon production is generated from dimension 8 operators through the following effective Lagrangian
\begin{equation}
\mathcal{L}_{\rm eff}\supset\frac{\sqrt{4\pi\alpha}}{M_{\rm int}^4}(\chi^\dagger\bar{\sigma}^\mu\chi)(e^\dagger\bar{\sigma}^\nu e)[a_LF_{\mu\nu}+b_L\tilde{F}_{\mu\nu}]+\{L\rightarrow R,e\rightarrow\bar{e}\}
\end{equation}
where two-component spinor notation \cite{Dreiner:2008tw} is used and the coupling constants $\{h_L,a_L,b_L,h_R,a_R,b_R\}$ should naturally be order one numbers.  The relevant cross-sections are $\langle\sigma_{\rm FSR}v\rangle\approx c_{\rm FSR}\alpha\frac{m_e^2}{M_{\rm int}^4}\ln\left(\frac{4m_\chi^2}{m_e^2}\right)$ and $\langle\sigma_{\rm eff}\rangle\approx c_{\rm eff}\alpha\frac{(2m_\chi)^6}{M_{\rm int}^8}$ where $c_{\rm FSR}\propto h_L^2+h_R^2$ and $c_{\rm eff}\propto a_L^2+\cdots$ include the appropriate order one coupling constants together with the $\pi$ factors from the phase space integration.

Finally, when scalar boson DM $\phi$ decays directly to one electron-positron pair, higher-order operators dominate for small $M_{\rm int}$ while FSR dominates for large $M_{\rm int}$.  Indeed, FSR is generated mainly by the dimension 5 operator
\begin{equation}
\mathcal{L}_{\rm FSR}\supset\frac{hm_e}{M_{\rm int}}\phi(\bar{e}e+e^\dagger\bar{e}^\dagger)
\end{equation}
while direct photon production is generated from dimension 7 operators through the following effective Lagrangian
\begin{equation}
\mathcal{L}_{\rm eff}\supset\frac{\sqrt{4\pi\alpha}}{M_{\rm int}^3}\partial^\mu\phi(e^\dagger\bar{\sigma}^\nu e)[a_LF_{\mu\nu}+b_L\tilde{F}_{\mu\nu}]+\{L\rightarrow R,e\rightarrow\bar{e}\}
\end{equation}
where two-component spinor notation \cite{Dreiner:2008tw} is used and the coupling constants
$\{h,a_L,b_L,a_R,b_R\}$ should naturally be order one numbers.
This operator has also been considered in Ref \cite{barger}.
The relevant decay rates are
$\Gamma_{\rm FSR}\approx c_{\rm FSR}\alpha\frac{m_e^2m_\phi}{M_{\rm int}^2}\ln\left(\frac{m_\phi^2}{m_e^2}\right)$
and $\Gamma_{\rm eff}\approx c_{\rm eff}\alpha\frac{m_\phi^7}{M_{\rm int}^6}$ where $c_{\rm FSR}\propto h^2$
and $c_{\rm eff}\propto a_L^2+\cdots$ include the appropriate order one coupling constants together
with the $\pi$ factors from the phase space integration.

For all scenarios both processes (final state radiation and direct photon production) have comparable contributions when
\begin{equation}\label{eqn:Mint}
M_{\rm int}^*\approx M\left(\frac{M}{m_e}\right)^{\frac{1}{2}}\left[\frac{h_L^2+h_R^2}{a_L^2+b_L^2+a_R^2+b_R^2}\ln\left(\frac{M^2}{m_e^2}\right)\right]^{-\frac{1}{4}}\approx10^6\,{\rm GeV}
\end{equation}
where we assumed order one coupling constants and $M\approx1$ TeV.  Therefore higher-order operators dominate over FSR for $M_{\rm int}\lesssim M_{\rm int}^*$ while FSR dominates over higher-order operators for $M_{\rm int}\gtrsim M_{\rm int}^*$.  Obviously for the scalar boson DM decay the decay rates are taken to be small enough such that DM is long-lived.  This is possible if the operator coefficients are small, which occurs for example when the effective dimension of the scalar boson DM is higher.  This is the case when the scalar boson DM is a composite field (like a glueball).  In this case, equation (\ref{eqn:Mint}) is still valid since the effective dimensions of the operators for both final state radiation and direct photon production increase together.  A more detailed analysis can be found in Appendix \ref{app:OPS}.

\subsection{$\mbox{DM}+\mbox{DM}\rightarrow2\phi$ and $\mbox{DM}\rightarrow2\phi$ followed by $\phi\rightarrow e^++e^-+\gamma$}\label{subsec:2ee1photon}

Here the computation of the photon DOS involves once more a convolution of the boosted (FSR or higher-order operators) photon DOS obtained in section \ref{subsec:1ee1photon} with the appropriate DOS of DM annihilation/decay to two bosons as described in Appendix \ref{app:DOS}.

\subsubsection*{Final state radiation of photons}

\begin{figure}[ht]
\begin{center}
\includegraphics[scale=1.2]{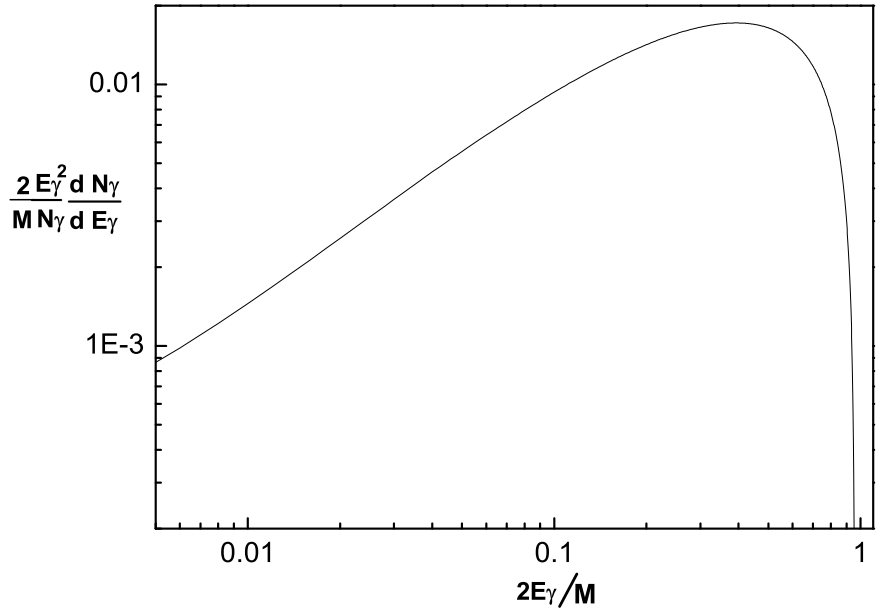}
\caption{Photon spectral distribution for $\mbox{DM}+\mbox{DM}\rightarrow2\phi$ and $\mbox{DM}\rightarrow2\phi$
followed by $\phi\rightarrow e^++e^-+\gamma$ from FSR with $M=2000$ GeV and $m_\phi=400$ GeV.
The distribution peaks at $2E_\gamma/M = E_{\gamma}/E_{\gamma}^{\rm max} \simeq 0.394$.}
\label{fig:BoostedFSR}
\end{center}
\end{figure}

Defining the boson mass as $m_\phi$, the photon DOS from FSR of a single photon from the electrons or the positrons is thus
\begin{equation}\label{eqn:BoostedFSR}
\frac{1}{N_{\gamma}}\frac{dN_\gamma}{dE_\gamma}\simeq\frac{\alpha}{\pi}\frac{1}{\sqrt{E_\phi^2-m_\phi^2}}\left\{
\begin{array}{ll}
f\left(\frac{2E_\gamma}{E_\phi-|\vec{p}_\phi|}\right)-f\left(\frac{2E_\gamma}{E_\phi+|\vec{p}_\phi|}\right) & \mbox{for $E_d<E_\gamma<E_\gamma^{{\rm int}}$}\\
f\left(1-\frac{4m_e^2}{m_\phi^2}\right)-f\left(\frac{2E_\gamma}{E_\phi+|\vec{p}_\phi|}\right) & \mbox{for $E_\gamma^{{\rm int}}<E_\gamma<E_\gamma^{{\rm max}}$}
\end{array}\right.
\end{equation}
where the function $f$ is
\begin{multline}
f(x)=\frac{x^2+x-2}{x}\ln\left(\frac{m_\phi^2(1-x)}{m_e^2}\right)-x+1\\
-2\left[1-\ln\left(\frac{m_e^2}{m_\phi^2}\right)\right]\ln(x)+2\,\mbox{dilog}(1-x)
\end{multline}
and the boson energy and momentum are $E_\phi=\frac{M}{2}$ and
$|\vec{p}_\phi|=\frac{M}{2}\sqrt{1-\frac{4m_\phi^2}{M^2}}$ respectively.
The limits on the photon energy are
\begin{equation*}
E_\gamma^{{\rm int}}=\frac{E_\phi-|\vec{p}_\phi|}{2}\left(1-\frac{4m_e^2}{m_\phi^2}\right)\hspace{0.5cm}\mbox{and}\hspace{0.5cm}E_\gamma^{{\rm max}}=\frac{E_\phi+|\vec{p}_\phi|}{2}\left(1-\frac{4m_e^2}{m_\phi^2}\right).
\end{equation*}
The photon spectrum is simply twice the photon DOS ($N_\gamma=2$) which is shown in figure \ref{fig:BoostedFSR} for a given boson mass.

Again it is possible to relate the FSR photon DOS of this subsection to the FSR photon DOS
of the previous subsection by letting the boson mass approach half the parameter $M$, i.e. $m_\phi=M/2$.
This corresponds to DM annihilation/decay to two bosons \textit{at rest} in the DM center of mass frame
which subsequently decay to one electron-positron pair each with a single FSR photon.
Notice that, in this limit, the $(E_\phi^2-m_\phi^2)^{-\frac{1}{2}}$ prefactor
in the photon DOS is very important for the matching to work.

\subsubsection*{Direct photon production from higher-order operators}

\begin{figure}[ht]
\begin{center}
\includegraphics[scale=1.2]{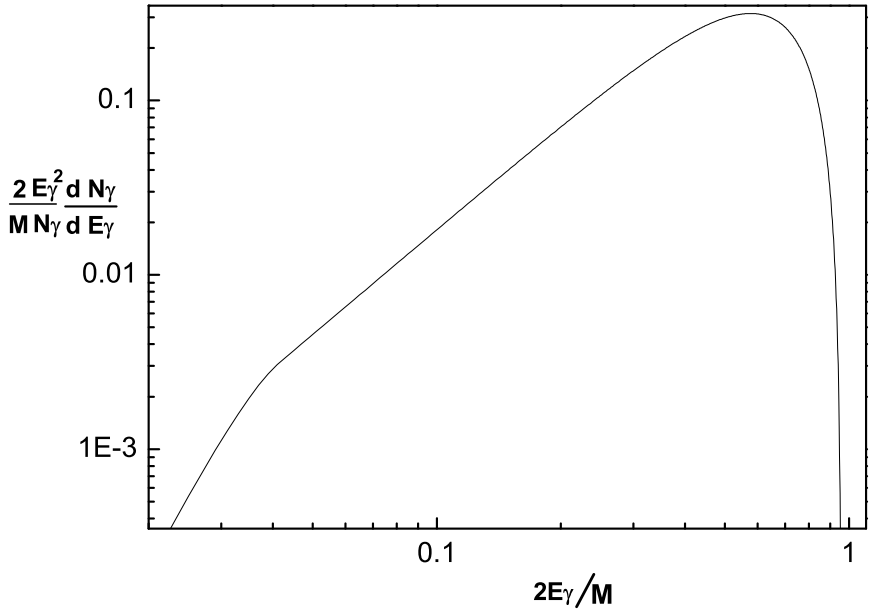}
\caption{Photon spectral distribution for $\mbox{DM}+\mbox{DM}\rightarrow2\phi$ and
$\mbox{DM}\rightarrow2\phi$ followed by $\phi\rightarrow e^++e^-+\gamma$
from the leading short range contact interaction with
$M=2000$ GeV and $m_\phi=400$ GeV, and neglecting the electron mass.
The distribution peaks at $2E_\gamma/M = E_{\gamma}/E_{\gamma}^{\rm max} \simeq 0.577$.}
\label{fig:Boostedoperator}
\end{center}
\end{figure}

As previously shown, photon production from scalar boson decay to
electron-positron pair is not dominated by FSR when the typical
scale of the interactions $M_{\rm int}$ is low.  In that case, the
photon DOS from higher-order operators is given by
\begin{equation}
\frac{1}{N_{\gamma}}\frac{dN_\gamma}{dE_\gamma}=\left\{
\begin{array}{ll}
\frac{160[2m_\phi^2(M^2-m_\phi^2)-3M(M^2-2m_\phi^2)E_\gamma]E_\gamma^3}{3m_\phi^8} & \mbox{for $0<E_\gamma<\frac{E_\phi-|\vec{p}_\phi|}{2}$}\\
\frac{5[(E_\phi+|\vec{p}_\phi|)^4-16(2E_\phi+2|\vec{p}_\phi|-3E_\gamma)E_\gamma^3]}{3|\vec{p}_\phi|(E_\phi+|\vec{p}_\phi|)^4} & \mbox{for $\frac{E_\phi-|\vec{p}_\phi|}{2}<E_\gamma<\frac{E_\phi+|\vec{p}_\phi|}{2}$}
\end{array}\right.
\end{equation}
where the boson energy and momentum are $E_\phi=\frac{M}{2}$ and
$|\vec{p}_\phi|=\frac{M}{2}\sqrt{1-\frac{4m_\phi^2}{M^2}}$
respectively.  The photon DOS, which is shown in figure
\ref{fig:Boostedoperator}, is canonically normalized and the photon
spectrum is simply twice the photon DOS ($N_\gamma=2$).  Once more,
it is possible to relate the photon DOS obtained in this subsection
with the photon DOS obtained in the previous subsection for
higher-order operators by taking the limit $m_\phi=\frac{M}{2}$.


\section{Photons from taus}\label{sec:taus}

Tau leptons represent another interesting decay mode for
dark matter annihilation/decay.
Since baryon number is conserved in tau decays, and
the tau mass is less than the sum of the proton and neutron
masses, $m_{\tau} < m_p + m_n$, tau decays include only
(anti-)leptons and (anti-)mesons, with no (anti-\nolinebreak[4])
baryons.
Dark matter that annihilates/decays preferentially to tau
leptons is therefore not necessarily in conflict with stringent
limits on the anti-proton flux in cosmic rays.
However, tau leptons do provide an interesting source of photons
since tau decays include a significant fraction of neutral
pions, $\tau \to X + \pi^0$, that subsequently decay to photons,
$\pi^0 \to 2 \gamma$.


The average number of $\pi^0$ produced in a single $\tau$ decay is approximately $\langle{N}_{\pi^0}\rangle \simeq 0.51$,
giving an average of roughly one photon per $\tau$ decay,
$N_{\gamma}  / N_{\tau} \simeq 1$.
Most $\pi^0$ from $\tau $ decays
come from the hadronic one-prong decay modes $\tau^-\rightarrow\nu_\tau+\rho^-\rightarrow\nu_\tau+\pi^-+\pi^0$
(branching fraction $25.4\%$) and $\tau^-\rightarrow\nu_\tau+a_1^-\rightarrow\nu_\tau+\pi^-+2\pi^0$ (branching fraction
approximately $9\%$  \cite{Graesser:2008qi}).
These two decay modes account for approximately $85\%$ of all the neutral pions arising from $\tau$ decay.
The other main sources of neutral pions are the three-prong mode
$\tau^-\rightarrow\nu_\tau+2\pi^-+\pi^++\pi^0$ (branching fraction
$4.3\%$) and the one-prong mode $\tau^-\rightarrow\nu_\tau+\pi^-+3\pi^0$ (branching fraction $1.1\%$),
as well as continuum contributions.
All branching fractions are taken from the PDG \cite{Amsler:2008zzb}.
Below we include only the dominant decays through the $\rho$ and $a_1$ resonances.

To obtain the photon DOS we first need to obtain the DOS of neutral pions.
We have explicitly computed the contribution to the pion spectrum from the principal
decay modes, with intermediate $\rho$ and $a_1$ vector meson resonances.
The general
$\tau$ differential decay rate takes the form
\begin{multline}
d\Gamma\propto\int d\Pi_2(\tau\rightarrow\nu_\tau+v)\,dm_v^2\,d\Pi_n(v\rightarrow n \pi)\\
\times|\check{\mathcal{M}}_{\mu,\pm}
  (\tau\rightarrow\nu_\tau+v)\mathcal{P}_v^{\mu\nu}(m_v^2)\hat{\mathcal{M}}_\nu(v\rightarrow n \pi)|^2
\end{multline}
The $\pi^0$ DOS is obtained by removing a single $\pi^0$ from the final state phase space integration.
Here $\mathcal{P}^{\mu\nu}_v(m_v^2)$ is the vector meson propagator,
$\check{\mathcal{M}}_{\mu,\pm}(\tau\rightarrow\nu_\tau+v)$ denotes
the matrix element for a $\tau$ of helicity $h=\pm{1 \over 2} $
corresponding to right or left handed respectively, to decay to a vector meson, and
$\hat{\mathcal{M}}_\nu(v\rightarrow n\pi)$ denotes
the matrix element for the vector meson to decay to $n$ pions.

First consider the decay through the $\rho$ resonance.
Obtaining the pion spectrum from the sequence of two-body
cascades $\tau\rightarrow\nu+\rho\rightarrow\nu+2\pi$ uses many of the techniques as used in the scalar
cascades in Section \ref{sec:direct}.
The major differences are first, nonconstant matrix elements,
resulting from the nonzero spin of the intermediate $\rho$, and second,
the large decay width of the $\rho$, which necessitates the use of a
Breit-Wigner with a running width.
We use a $\rho$ mass and width of
$m_{0,\rho}=770\,{\rm MeV}$ and
$\Gamma_{0,\rho}=150\,{\rm MeV}$.
Next consider the decay through the $a_1$ resonance which differs from the $\rho$ mode in that the final
decay $a_1\rightarrow3\pi$ is not two-body.
The spectrum of the observed pion therefore requires additional integrations
over the phase space of the unobserved pions in the final state.
Following \cite{Graesser:2008qi,Bullock:1992yt}, we use relatively
simple parameterizations given in \cite{Kuhn:1990ad}
for both the $a_1\rightarrow3\pi$ matrix element
and running width.
We use an $a_1$ mass and width of
$m_{0,a}=1.22\,{\rm GeV}$ and $\Gamma_{0,a}=420\,{\rm MeV}$.

\begin{figure}[ht]
\begin{center}
\includegraphics[scale=1.2]{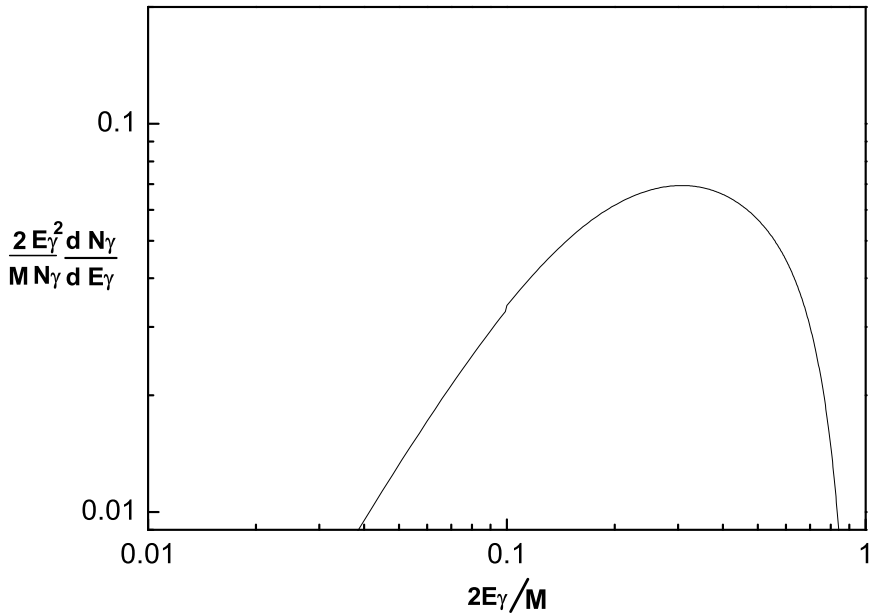}
\caption{
Photon spectral distribution for
$\mbox{DM}+\mbox{DM}\rightarrow 2 \tau$ and
$\mbox{DM}\rightarrow 2 \tau $ followed by
$\tau \rightarrow X+ \pi^0$ and $\pi^0 \to 2 \gamma$
in the limit $M \gg m_\tau$.
The distribution peaks at
$2E_\gamma/M = E_{\gamma}/E_{\gamma}^{\rm max} \simeq 0.307$.}
\label{fig:Tau}
\end{center}
\end{figure}

The photon DOS may be obtained from the pion DOS by convolution, as before.
We work in the collinear limit $M \gg m_{\tau}$ in which the
components of the $\pi^0$, and therefore photon,
momentum transverse to the tau direction of
motion in the origin annihilation/decay frame are irrelevant.
The results for the normalized photon DOS under the
assumption that the DM annihilates/decay to a single
 tau--anti-tau pair are well fit by the parameterized functional
form
\beq
{ E_{\gamma}^{\rm max} \over N_{\gamma} }
 { d N_{\gamma} \over d E_{\gamma} }  =
f(E_{\gamma} / E_{\gamma}^{\rm max} ) e^{-g(E_{\gamma} / E_{\gamma}^{\rm max} )}
\eq
where
$E_{\gamma}^{\rm max} = M/2$,
and
\begin{align}
f(x) &= x^a \sum_{n=0}^{4} b_n x^n \nonumber \\
g(x) &= \sum_{n=1}^{3} c_n x^n
\end{align}
For positive or right handed helicity the best fit parameters are
$a_{+} = -0.192$,
$b_{+,i} = \{ 3.90, -2.26, -15.59, 32.96, -19.45 \}$,
$c_{+,i} = \{ 8.94, -16.10, -19.45 \} $,
while for negative or left handed helicity the best fit parameters
are
$a_{-} = -0.040$,
$b_{-,i} = \{ 5.55, 4.29, -10.94, 0 , 0 \}$,
$c_{-,i} = \{ 7.36 ,0 ,0 \} $.
The differences between the photon DOS arising
from right- and left-handed taus are only very minor.
It seems unlikely that a measurement of the photon spectrum alone could ever
distinguish the helicity of tau's arising from DM annihilation/decay.
The photon distribution arising from DM annihilation/decay directly
to a single tau--anti-tau pair and averaging over tau helicity
is shown in Fig. \ref{fig:Tau}.
In this case the photon spectrum is roughly twice the photon DOS, $N_\gamma \simeq 2$,
since there is roughly one photon per tau on average and two tau's per DM annihilation/decay.

\begin{figure}[ht]
\begin{center}
\includegraphics[scale=1.2]{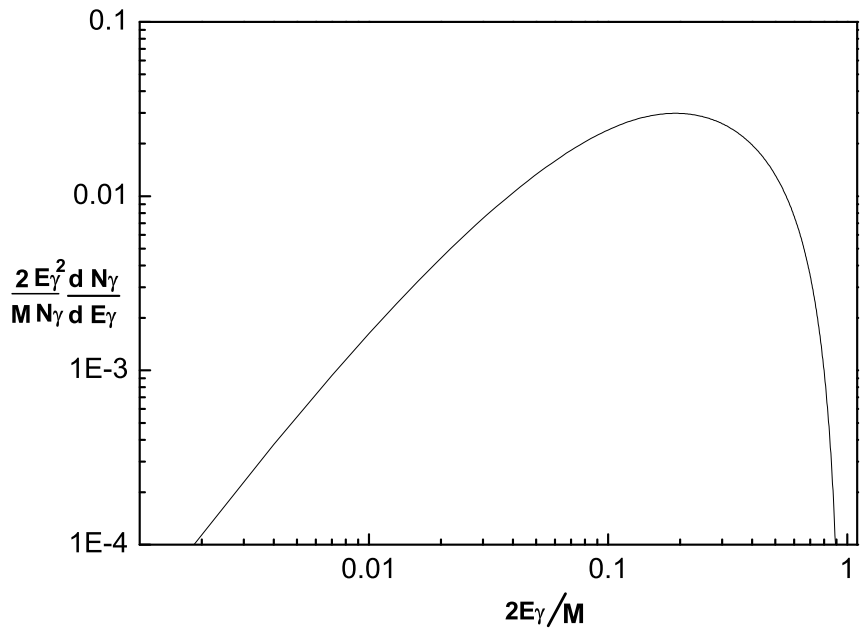}
\caption{
Photon spectral distribution for
$\mbox{DM}+\mbox{DM}\rightarrow 2 \phi$ and
$\mbox{DM}\rightarrow 2 \phi $ followed by
$\phi \to 2 \tau$ and
$\tau \rightarrow X+ \pi^0$ and $\pi^0 \to 2 \gamma$
in the limit $M \gg m_{\phi} \gg m_\tau$.
The distribution peaks at $2E_\gamma/M \simeq 0.192$.}
\label{fig:Boostedtau}
\end{center}
\end{figure}

If we assume DM particles annihilate/decay to two intermediate bosons $\phi$, each of which
subsequently decays to a tau--anti-tau pair, then we again need to boost the DOS
obtained above and convolve it with the appropriate DOS of DM annihilation/decay
to two bosons, as explained in Appendix \ref{app:DOS}.
The final photon spectral distribution for this case
in the collinear limits $M \gg m_{\phi} \gg m_{\tau}$,
and averaging over tau helicities
is shown in Fig. \ref{fig:Boostedtau}.
The photon spectrum in this case is roughly four times the photon DOS, $N_\gamma \simeq 4$.
We can see that the final photon DOS is softer compared to Fig. \ref{fig:Tau} due to extra intermediate
state.


\section{Photon spectra and flux}\label{sec:photon}

\subsection{Photon spectra}\label{subsec:spectra}

\begin{figure}[ht]
\begin{center}
\includegraphics[scale=1.2]{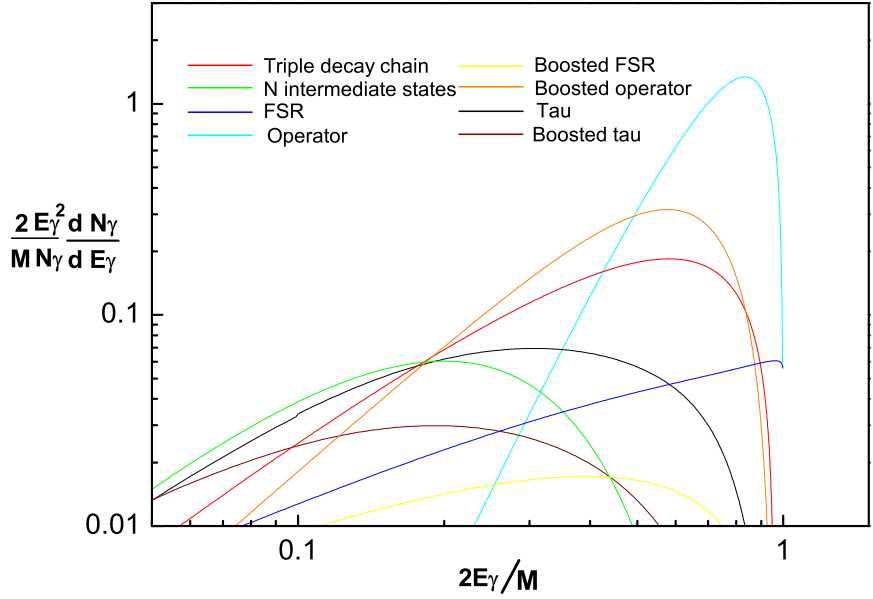}
\caption{Photon spectral disbributions arising from different DM
annihilation/decay scenarios for $M=2000$ GeV, $m_\phi=400$ GeV and
$m_\pi=0.14$ GeV.
For DM annihilation/decay to $N$ intermediate
$\phi$ bosons the limit $M \gg m_\phi$ is presented with $N=10$.
For DM annihilation/decay through intermediate $\tau$'s
the limits $M \gg m_\tau$ or $M \gg m_\phi \gg m_\tau$ are presented.
In each case the photon DOS is normalized to unit probability,
except for FSR and Boosted FSR which are normalized with respect to the leading decay without FSR.}
\label{fig:All}
\end{center}
\end{figure}

Strickly speaking, everything we have computed up to now are photon DOS in different scenarios.  The different photon DOS are superimposed in figure \ref{fig:All}.
However, in the likely case where several scenarios add up, the different spectra must be added with the appropriate weight.
The full photon spectrum obtained is given by
\begin{equation}
\frac{dN_\gamma^{{\rm total}}}{dE_\gamma}=\sum_{i}\mbox{Br}(i)\frac{dN_\gamma^{(i)}}{dE_\gamma}
\end{equation}
where the sum is over all allowed scenarios $i$, with photon spectrum $\frac{dN_\gamma^{(i)}}{dE_\gamma}$ and branching ratio $\mbox{Br}(i)$.
In comparing with the background of gamma ray photons, the only difference between DM annihilation and DM decay
comes from the power of the DM density profile in the appropriate equations as discussed in the next subsection.


\subsection{Photon flux}\label{subsec:los}

An observation of gamma rays from dark matter annihilation/decay involves
not only the spectrum but also the absolute magnitude.
The photon flux
\beq
\Phi_{\gamma} \equiv { d N_{\gamma} \over dA~dt~d \Omega}
\eq
where $dA$ and $d \Omega$ are the detector area and solid angle elements,
can be computed from an integral over the source along the line of sight
(los) \cite{Serpico:2009vz} since gamma rays are not significantly attenuated
on galactic length scales.
For annihilation of a single species of DM particle that is its own anti-particle
the spectral flux is given by
\begin{equation}
\label{eqn:losann}
{ d \Phi_\gamma \over d E_{\gamma} } (E_\gamma,\Omega)=\frac{\langle\sigma
v\rangle}{4\pi m_{\rm DM}^2}\frac{dN_\gamma}{dE_\gamma}\int_{\rm los}\rho_{\rm DM}^2(r,\Omega)dr
\end{equation}
where $\langle\sigma v\rangle$ is the phase space averaged annihilation cross section times velocity,
$\rho_{\rm DM}$ is the dark matter density, and $d N_{\gamma} / d E_\gamma$ is the photon spectrum
with $N_\gamma = \int dE_\gamma (d N_{\gamma} / d E_\gamma)$ photons emitted per annihilation.
If the DM is composed of distinct particle and anti-particle particle species with equal densities
that can annihilate only through the particle--anti-particle channel, the flux
(\ref{eqn:losann}) should be multiplied by an additional factor of ${1 \over 4} $.
In terms of dimensional units the annihilation spectral flux (\ref{eqn:losann}) may be written
\begin{equation}
\frac{d \Phi_\gamma(E_\gamma,\Omega) / d E_{\gamma} }{{\rm cm}^{-2}\cdot{\rm s}^{-1}\cdot{\rm sr}^{-1}}
  \simeq
5.6 \times 10^{-10}\left(\frac{\langle\sigma
v\rangle}{3\times 10^{-26}\,{\rm cm}^3\cdot{\rm s}^{-1}}\right)\left(\frac{100\,{\rm GeV}}{m_{\rm DM}}\right)^2\frac{dN_\gamma}{dE_\gamma}
J_2(\Omega)
\end{equation}
where
\begin{equation}
J_2(\Omega) \equiv\frac{1}{8.5\,{\rm kpc}}\left(\frac{1}{0.3\,{\rm GeV}\cdot{\rm cm}^{-3}}\right)^2\int_{\rm los}\rho^2(r,\Omega)dr
\end{equation}
is a dimensionless order one factor that represents
astrophysical parameters.

For decay of a single species of DM
the spectral flux is given by
\begin{equation}\label{eqn:losdecay}
{ d \Phi_\gamma \over d E_{\gamma} } (E_\gamma,\Omega)=\frac{\Gamma}{4\pi
m_{\rm DM}}\frac{dN_\gamma}{dE_\gamma}\int_{\rm los}\rho_{\rm DM}(r,\Omega)dr
\end{equation}
where $\Gamma$ is the DM decay rate with
$N_\gamma = \int dE_\gamma (d N_{\gamma} / d E_\gamma)$ photons emitted per decay.
In terms of dimensional units the decay spectral flux may be written
\begin{equation}
\frac{d \Phi_\gamma(E_\gamma,\Omega)/ d E_{\gamma} }{{\rm cm}^{-2}\cdot{\rm s}^{-1}\cdot{\rm sr}^{-1}}\approx
6.26 \times 10^{-9}
  \left(\frac{\Gamma}{10^{-27}\,{\rm s}^{-1}}\right)\left(\frac{100\,{\rm GeV}}{m_{\rm DM}}\right)\frac{dN_\gamma}{dE_\gamma}
  J_1(\Omega)
\end{equation}
where
\begin{equation}
J_1(\Omega) \equiv
  \frac{1}{8.5\,{\rm kpc}}\frac{1}{0.3\,{\rm GeV}\cdot{\rm cm}^{-3}}\int_{\rm los}\rho_{\rm DM}(r,\Omega)dr
\end{equation}
is a dimensionless order one  factor characterizing the astrophysical parameters.
With the spectral flux equations (\ref{eqn:losann}) and (\ref{eqn:losdecay}) comparisons
with the gamma ray background are straightforward.


\section{Conclusion}\label{sec:conc}

High-energy photons are hardly deflected when they propagate through the galaxy.  This is an advantage over charged particles, like electrons and positrons, which interact with the galactic magnetic field.  Moreover, once charged particles are accelerated, photons are always produced due to radiation.  The different photon spectra one can obtained are good candidates to provide information on the process(es) generating the photons.  In this paper, we studied the spectra of high-energy photons generated by various dark matter annihilation/decay scenarios including: direct photon production from arbitrary two-body decay chains, final state radiation from charged particles generated by dark matter annihilation/decay, direct photons production together with charged particles from higher-order operators, and the special case where photons are produced from taus generated by dark matter annihilation/decay.

We noted also that, for processes which generate photons and leptons, effective field theory allows a comparison between the spectra from final state radiation and the ones from direct photon production due to higher-order operators.  Interestingly, we found that, except for scalar dark matter annihilation or decay and Majorana dark matter annihilation, all the other dark matter scenarios are dominated by final state radiation.  Moreover, for these exceptions (scalar boson dark matter annihilation and decay and Majorana dark matter annihilation), with a dark matter mass of $\mathcal{O}(1\,{\rm TeV})$ we found that direct photon production from higher-order operators dominates if the scale of the leading operator is lower than $\mathcal{O}(1000\,{\rm TeV})$.  Otherwise final state radiation still dominates.  Finally, it is also very interesting to see that the hardest spectrum among the spectra studied here (see figure \ref{fig:All}) comes from these exceptions, i.e. direct photon production from higher-order operators.

Once the flux of cosmic gamma rays is measured, an eventual dark matter signal could be compared with the different spectra presented here and a great deal of information on the nature of dark matter at the particle physics level could be deduced.

\section*{Acknowledgments}

This research was supported by DOE grant DE-FG02-96ER40949.
We would like to thank IAS for their hospitality during the ``Current Trends in Dark Matter'' workshop,
and Rouven Essig for comments on this manuscript.


\appendix

\section{Density of states}\label{app:DOS}

In this appendix we review how the DOS is obtained from general
considerations.  The DOS for $\Phi$ annihilation/decay to $\phi$ is
simply given by
\begin{equation}
\frac{1}{N}\frac{dN}{dE}=\frac{1}{\langle\sigma
v\rangle}\frac{d\langle\sigma
v\rangle}{dE}\hspace{0.5cm}\mbox{and}\hspace{0.5cm}\frac{1}{N}\frac{dN}{dE}=\frac{1}{\Gamma}\frac{d\Gamma}{dE}
\end{equation}
respectively, and is thus normalized to one.  For example, assuming constant averaged matrix element squared, the DOS in the $\Phi$ center of mass frame for $2\rightarrow2$ annihilation in the s-wave approximation and $1\rightarrow2$ decay is simply obtained from the phase space,
\begin{multline}
\frac{1}{N}\frac{dN}{dE}=\left[\int\frac{d^3p}{EdE}\frac{d^3q}{E'}\delta^{(4)}(P-p-q)\right]\\
\times\left[\int\frac{d^3p}{E}\frac{d^3q}{E'}\delta^4(M-p-q)\right]^{-1}=\delta\left(E-\frac{M}{2}\right)
\end{multline}
where $P^\mu=(M,\vec{0})$ with $M=2m_\Phi$ for $\Phi$ annihilation and $M=m_\Phi$ for $\Phi$ decay.

For $\Phi$ decay in a boosted frame, the DOS follows from Lorentz
covariance.  Indeed $N\equiv\int dE\frac{dN}{dE}$ is a Lorentz
scalar thus $N^{{\rm CM}}=N^{{\rm Boost}}$.  Therefore one can
rewrite the boosted DOS as
\begin{equation*}
\frac{1}{N}\frac{dN}{dE}^{{\rm Boost}}=\int dE^{{\rm CM}}dz^{{\rm
CM}}\frac{dN^{{\rm CM}}}{N dz^{{\rm CM}}dE^{{\rm CM}}}=\int
dEdz^{{\rm CM}}\left|\frac{\partial(E^{{\rm CM}},z^{{\rm
CM}})}{\partial(E,z^{{\rm CM}})}\right|\frac{dN^{{\rm CM}}}{N
dz^{{\rm CM}}dE^{{\rm CM}}}
\end{equation*}
or
\begin{equation}
\frac{1}{N}\frac{dN^{{\rm Boost}}}{dE}=\int dz^{{\rm
CM}}\left|\frac{\partial(E^{{\rm CM}},z^{{\rm
CM}})}{\partial(E,z^{{\rm CM}})}\right|\frac{dN^{{\rm CM}}}{N
dz^{{\rm CM}}dE^{{\rm CM}}}.
\end{equation}
Here $z^{{\rm CM}}=\cos\theta=\hat{p}_\Phi\cdot\hat{p}_\phi^{{\rm
CM}}$ is the angle between the boosted $\Phi$ and the unboosted
$\phi$.  Since the DOS in the $\Phi$ rest frame is uniform, then
$\frac{dN^{{\rm CM}}}{N dz^{{\rm CM}}dE^{{\rm
CM}}}=\frac{1}{2}\frac{dN^{{\rm CM}}}{N dE^{{\rm CM}}}$.  Finally,
the $\phi$ energy in the $\Phi$ center of mass frame is related to
the $\phi$ energy in the boosted frame by
\begin{equation}
E=\frac{E_\Phi}{m_\Phi}\left(E^{{\rm CM}}+z^{{\rm CM}}\sqrt{{E^{{\rm CM}}}^2-m_\phi^2}\sqrt{E_\Phi^2-m_\Phi^2}\right).
\end{equation}
The bounds on $z$ can be found from the bounds in the $\Phi$ center
of mass frame plus the physical constraint that $-1<z<1$.  Boosting
the DOS of the previous example in the frame where $\Phi$ has
four-momentum
$p_\Phi^\mu=E_\Phi(1,\hat{p}_\Phi\sqrt{1-m_\Phi^2/E_\Phi^2})$, one
gets
\begin{equation}
\frac{1}{N}\frac{dN^{{\rm
Boost}}}{dE}=\frac{1}{\sqrt{1-4m_\phi^2/m_\Phi^2}}\frac{1}{\sqrt{E_\Phi^2-m_\Phi^2}}
\end{equation}
where the $\phi$ energy is bounded by
\begin{eqnarray*}
E^{{\rm max}} &=& \frac{E_\Phi}{2}\left[1+\sqrt{1-4m_\phi^2/m_\Phi^2}\sqrt{1-m_\Phi^2/E_\Phi^2}\right]\\
E^{{\rm min}} &=& \frac{E_\Phi}{2}\left[1-\sqrt{1-4m_\phi^2/m_\Phi^2}\sqrt{1-m_\Phi^2/E_\Phi^2}\right].
\end{eqnarray*}

\subsection*{Convolution and matching of DOS}

For a two-body decay chain $\phi_0\rightarrow2\phi_1$ to
$\phi_{k-1}\rightarrow2\phi_k$ the DOS in the $\phi_0$ rest frame
can be found by iteration and is given by
\begin{equation}
\frac{dN_k}{NdE_k}=\int dE_{k-1}\cdots dE_1\frac{dN_1^{{\rm
CM}}}{N_1 dE_1}\frac{dN_2^{{\rm Boost}}}{N_2
dE_2}\cdots\frac{dN_k^{{\rm Boost}}}{N_k dE_k}=\int
dE_{k-1}\frac{dN_{k-1}}{N_{k-1} dE_{k-1}}\frac{dN_k^{{\rm
Boost}}}{N_{k} dE_k}
\end{equation}
where the bounds are complicated functions of the energies.  For
example, with the same assumptions as above, the two-body decay
chain $\phi_0\rightarrow2\phi_1$ followed by
$\phi_1\rightarrow2\phi_2$ gives
\begin{eqnarray}
\frac{dN_2}{N_2 dE_2} &=& \int dE_1\frac{dN_1}{N_1 dE_1}\frac{dN_2^{{\rm Boost}}}{N_2 dE_2}\nonumber\\
 &=& \int dE_1\delta\left(E_1-\frac{m_0}{2}\right)\frac{1}{\sqrt{1-4m_2^2/m_1^2}}\frac{1}{\sqrt{E_1^2-m_1^2}}\nonumber\\
 &=& \frac{2}{m_0\sqrt{1-4m_1^2/m_0^2}\sqrt{1-4m_2^2/m_1^2}}
\end{eqnarray}
where the $\phi_2$ energy is bounded by
\begin{eqnarray*}
E_2^{{\rm max}} &=& \frac{m_0}{4}\left[1+\sqrt{1-4m_1^2/m_0^2}\sqrt{1-4m_2^2/m_1^2}\right]\\
E_2^{{\rm min}} &=& \frac{m_0}{4}\left[1-\sqrt{1-4m_1^2/m_0^2}\sqrt{1-4m_2^2/m_1^2}\right].
\end{eqnarray*}

By taking limits where intermediate particles are created \textit{at rest} in the center of mass frame of the parent particle, it is possible to match different DOS.  Indeed in the limit where $m_{i+1}\rightarrow\frac{m_i}{2}$, the decay chain is effectively cut by one step, with all the  previous steps being unchanged, the $i$-th step being deleted and the subsequent steps being modified due to the energy redistribution.  The DOS satisfies
\begin{multline}
\left[\lim_{m_{i+1}\rightarrow\frac{m_i}{2}}\frac{dN_k}{N_k dE_k}\right]_{m_i=2m_{i+1}}\\
=\left[\frac{dN_{k-1}}{N_{k-1}dE_{k-1}}\right]_{m_0=\frac{m_0}{2},\ldots,m_{i-1}=\frac{m_{i-1}}{2};m_i=m_{i+1},\ldots,m_{k-1}=m_k}
\end{multline}
where the $i$th step is not included on the LHS.  This allows us to
check the different DOS formula.


\section{FSR collinear divergence}\label{app:FSR}

In the calculation of FSR, one would have to deal with the collinear
divergence, which will show up when we take the massless limit for
$e^+ e^-$. Taking the electron mass into account, the three-body
decay rate does not suffer from a collinear divergence and it can be
easily computed using Dalitz coordinates \cite{Bergstrom:1989jr}.
Here we show the result for $\phi\rightarrow e^+ +e^-+\gamma$, where
FSR coming from the following vertex $\frac{h
m_e}{N_{int}}\phi(\bar{e}e+\bar{e^{\dag}}e^{\dag})$, one can get
spectrum as
\begin{multline}\label{eqn:FSRscalarDMdecay}
\frac{d\Gamma}{dE_\gamma}=\frac{\alpha h^2m_e^2}{8\pi^2M_{\rm int}^2}\left(\frac{m_\phi^2+(m_\phi-2E_\gamma)^2}{m_\phi E_\gamma}\ln\left[\frac{m_\phi(m_\phi-2E_\gamma)}{m_e^2}\right]\right.\\
\left.+\frac{2(m_\phi-2E_\gamma)}{E_\gamma}+\mathcal{O}(m_e^2)\right).
\end{multline}

Such spectrum has soft photon divergence, so we choose to normalize
the spectrum respect to its zero$th$ order approximation on
$\alpha$, which corresponds to the process $\phi\rightarrow e^+ +
e^-$, i.e.
\begin{equation}
\frac{dN_\gamma}{N_\gamma
dE_\gamma}\simeq\frac{1}{\Gamma_{\phi\rightarrow
e^++e^-}}\frac{d\Gamma}{dE_\gamma}.
\end{equation}
where $\Gamma_{\phi\rightarrow e^++e^-}=\frac{h^2m_e^2m_\phi}{8\pi
M_{\rm int}^2}$.

 The first piece of the formula is from long
distance contribution since this part is not sensitive to the
details of the vertex, and it will be logarithmically diverge when
we take $m_e$ goes to zero.  The second term is the finite piece
when we take massless limit of $e^+ e^-$.  This is the extra piece
comparing to collinear approximation where we only keep long
distance piece. In the collinear limit (see equation
(\ref{eqn:collinear})), after the last term of equation
(\ref{eqn:FSRscalarDMdecay}) is dropped, one gets the photon DOS
equation (\ref{eqn:FSR}) introduced in section \ref{sec:irr}.

Next, we consider the case where FSR gets one step of boosting, i.e.
DM annihilate/decay to intermediate particle $\phi$, then $\phi$
decays to $e^+ e^-$ with photon from FSR.

In a frame where $\phi$ has four-momentum
$p_\phi^\mu=E_\phi(1,\hat{p}_\phi\sqrt{1-m_\phi^2/E_\phi^2})$, the
boosted DOS can be found from the general formula.  However, since
the photon is massless, the Jacobian simplifies greatly, leading to
\begin{equation}
\frac{dN_\gamma^{{\rm Boost}}}{N_\gamma
dE_\gamma}=\int\frac{dz^{{\rm CM}}}{2}\frac{m_\phi}{E_\phi+z^{{\rm
CM}}\sqrt{E_\phi^2-m_\phi^2}}\frac{dN_\gamma^{{\rm CM}}}{N_\gamma
dE_\gamma^{{\rm CM}}}
\end{equation}
The integral is easily done using the new variable $w=\frac{m_\phi}{E_\phi+z^{{\rm CM}}\sqrt{E_\phi^2-m_\phi^2}}$, in terms of which $E_\gamma^{{\rm CM}}=wE_\gamma$.  The bounds on $w$ can be found from the bounds on $E_\gamma^{{\rm CM}}$ due to $E_\gamma^{{\rm CM}}(E_\gamma)$ and from the physical constraint $-1<z<1$, giving two different regimes,
\begin{equation}
\frac{m_\phi}{E_\phi+\sqrt{E_\phi^2-m_\phi^2}}<w<\min\left\{\frac{m_\phi}{E_\phi+\sqrt{E_\phi^2-m_\phi^2}},\frac{m_\phi}{2E_\gamma}\left(1-\frac{4m_e^2}{m_\phi^2}\right)\right\}
\end{equation}
as discussed in the text.  Convoluting this boosted DOS with the two-body decay DOS one obtains the DOS for $\mbox{DM}\rightarrow2\phi$ followed by $\phi\rightarrow e^++e^-+\gamma$ mentioned above.

\section{Higher-order operators}\label{app:OPS}

In some specific scenarios, FSR suffers a large chiral suppression due to the small electron mass.  In these cases, direct photon production from higher-order operators might dominate.  In this Appendix we determine the necessary conditions for which direct photon production from higher-order operators dominates over FSR.  To reach the most general conclusions, effective field theory is used throughout the analysis.  Therefore, the chirality rule, which states that chirality-violating operators must come with an overall mass term, is enforced.  Moreover, to evaluate the appropriate FSR annihilation cross-section/decay rate, the collinear limit is taken \cite{Birkedal:2005ep},
\begin{multline}\label{eqn:collinear}
\frac{dX_{e^++e^-+\gamma}}{dx}\simeq\frac{\alpha}{\pi}\left(\frac{1+(1-x)^2}{x}\ln\left[\frac{M^2(1-x)}{m_e^2}\right]\right)X_{e^++e^-}\\
\Rightarrow X_{\rm FSR}\approx\frac{\alpha}{\pi}X_{e^++e^-}\ln\left[\frac{M^2}{m_e^2}\right]
\end{multline}
where $x=2E_\gamma/M$ and $X=\langle\sigma v\rangle$ with $M=2m_{\rm DM}$ for DM annihilation or $X=\Gamma$ with $M=m_{\rm DM}$ for DM decay respectively.  Notice that for DM annihilation, DM is assumed to be almost at rest, thus the s-wave approximation can be taken.  The lowest-dimension electron operators are given in table \ref{tab:opsElectron}.  When the operator does not violate chirality, only the operators with $e$ alone are considered since operators mixing $e$ and $\bar{e}$ will lead to chirality-suppressed mixing terms.
\begin{table}[ht]
\begin{center}
\includegraphics[scale=0.65]{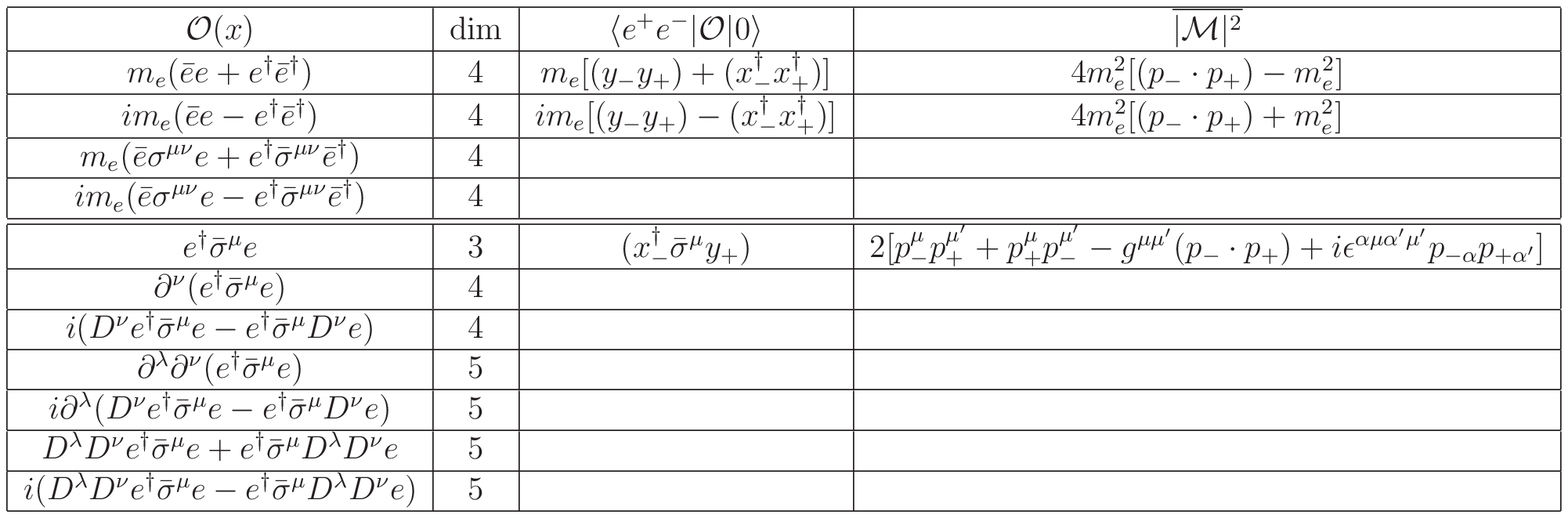}
\caption{Relevant electron-positron operators in the effective Lagrangian approach.  The empty boxes correspond to operators which are not needed in the analysis.}
\label{tab:opsElectron}
\end{center}
\end{table}
The photon can either appear in a covariant derivative, which is already taken into account in the electron operators, or in the field strength tensor as in table \ref{tab:opsPhoton}.
\begin{table}[ht]
\begin{center}
\includegraphics[scale=0.9]{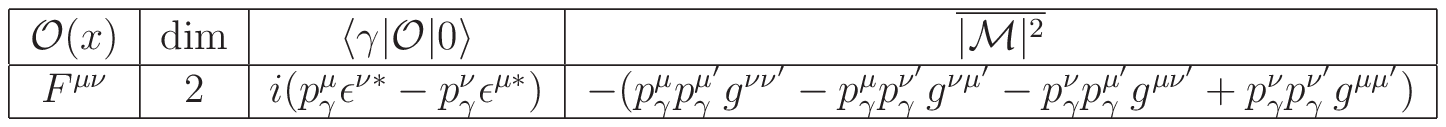}
\caption{Relevant photon operators in the effective Lagrangian approach.}
\label{tab:opsPhoton}
\end{center}
\end{table}
To simplify the analysis, the Lorentz indices are kept free and are contracted with the appropriate tensors ($g_{\mu\nu}$ or $\epsilon_{\mu\nu\lambda\rho}$) only at the end of the analysis.  In this way, one does not have to deal with the complete set of operators at this stage of the analysis (for example, the operator corresponding to the dual field strength tensor can be forgotten).

\subsection*{Scalar boson DM annihilation to $e^++e^-+\gamma$}

The relevant lowest-dimension scalar boson operators are given in table \ref{tab:opsScalarDMann}.
\begin{table}[ht]
\begin{center}
\includegraphics[scale=0.95]{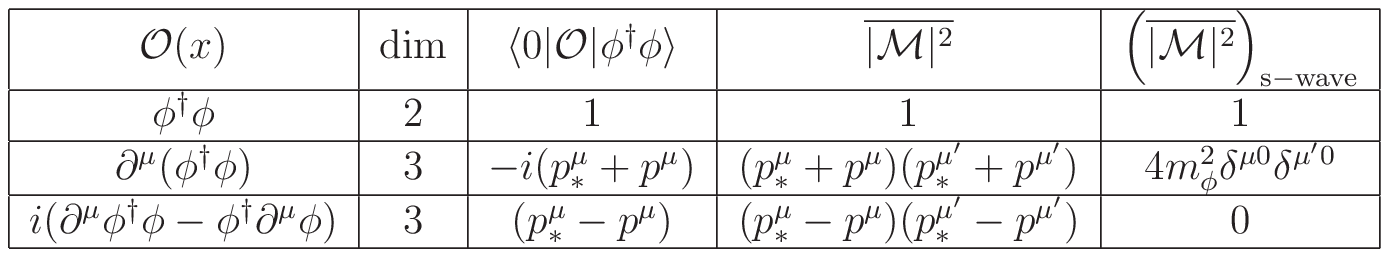}
\caption{Relevant scalar boson DM operators for DM annihilation.}
\label{tab:opsScalarDMann}
\end{center}
\end{table}
Combining operators of table \ref{tab:opsScalarDMann} with operators of table \ref{tab:opsElectron}, the leading operator for FSR seems to be constructed from $\partial^\mu(\phi^\dagger\phi)$ and $(e^\dagger\bar{\sigma}_\mu e)$.  It is a dimension 6 operator and it is not explicitly chirality-suppressed.  However, in the s-wave approximation, $p_+^\mu=(m_\phi,\vec{p}_e)$ and $p_-^\mu=(m_\phi,-\vec{p}_e)$, thus $\overline{|\langle e^+e^-|(e^\dagger\bar{\sigma}^0e)|0\rangle|^2}=2m_e^2$ which is chirality-suppressed.  Therefore, the leading FSR operators are dimension 6 but are chirality-suppressed.  For example, one such operator can be constructed out of $\phi^\dagger\phi$ and $m_e(\bar{e}e+e^\dagger\bar{e}^\dagger)$.  The related FSR cross-section is
\begin{equation}
\langle\sigma_{\rm FSR}v\rangle\approx c_{\rm FSR}\alpha\frac{m_e^2}{M_{\rm int}^4}\ln\left[\frac{4m_\phi^2}{m_e^2}\right]
\end{equation}
where $c_{\rm FSR}$ includes the operator coupling constant and the $\pi$ factors from the phase space integration.

For direct photon production from higher-order operators to dominate, the operators should not be chirality-suppressed and/or should be of lower dimension.  Combining again operators of table \ref{tab:opsScalarDMann} with operators of tables \ref{tab:opsElectron} and \ref{tab:opsPhoton} such that one photon can be created in the final state, the leading operators for direct photon production are dimension 8 and are not chirality-suppressed.  The operator made out of $\partial^\mu(\phi^\dagger\phi)$, $(e^\dagger\bar{\sigma}_\nu e)$ and $F_{\mu\nu}$ is an example.  Following the general rules of effective field theory, a factor of $\alpha$ should be put in front of this type of operators.  The related direct photon production cross-section is
\begin{equation}
\langle\sigma_{\rm eff}v\rangle\approx c_{\rm eff}\alpha\frac{(2m_\phi)^6}{M_{\rm int}^8}
\end{equation}
where $c_{\rm eff}$ includes the operator coupling constant and the $\pi$ factors from the phase space integration\footnote{$c_{\rm FSR}$ and $c_{\rm eff}$ have the same phase space factors since both are obtained from a three-particle final state.}.

It is now straightforward to compare FSR and direct photon production cross-sections,
\begin{equation}\label{eqn:FSRvsOPS}
\langle\sigma_{\rm FSR}v\rangle\approx\langle\sigma_{\rm eff}v\rangle\frac{c_{\rm FSR}}{c_{\rm eff}}\frac{m_e^2M_{\rm int}^4}{(2m_\phi)^6}\ln\left[\frac{4m_\phi^2}{m_e^2}\right]
\end{equation}
which leads to equation (\ref{eqn:Mint}) of section \ref{sec:irr}.  Direct photon production from higher-order operators will therefore dominate over FSR when $M_{\rm int}\lesssim M_{\rm int}^*$ (see equation (\ref{eqn:Mint})).  In the remaining subsections, the analysis is more concise since it closely follows what has been done here.

\subsection*{Majorana DM annihilation to $e^++e^-+\gamma$}

The relevant lowest-dimension Majorana operators are given in table \ref{tab:opsMajDMann}.
\begin{table}[ht]
\begin{center}
\includegraphics[scale=0.65]{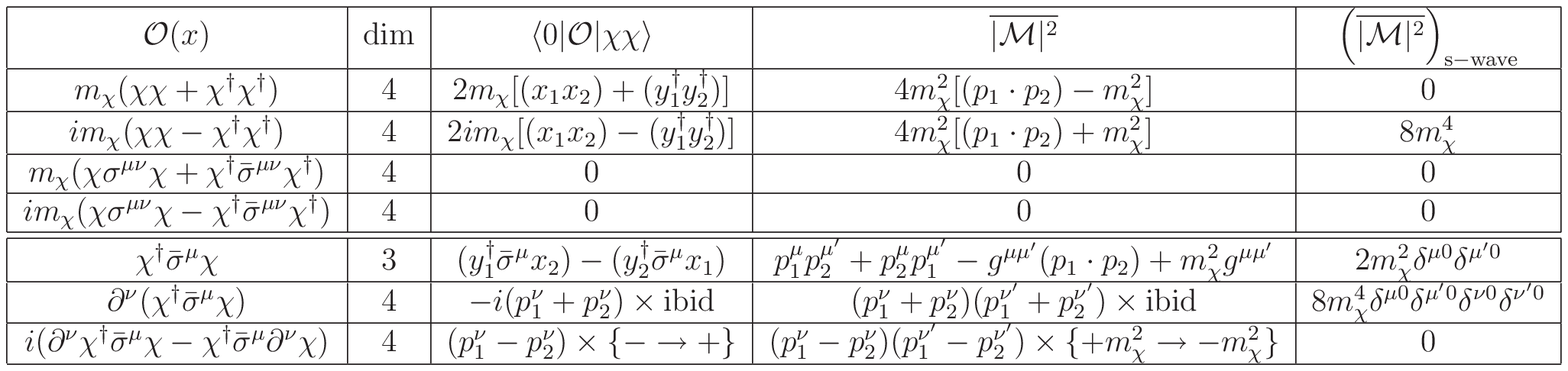}
\caption{Relevant Majorana DM operators for DM annihilation.  Notice that $\chi\bar{\sigma}^{\mu\nu}\chi=0$ since $\chi$ is a Majorana fermion.}
\label{tab:opsMajDMann}
\end{center}
\end{table}
Combining operators of table \ref{tab:opsMajDMann} with operators of table \ref{tab:opsElectron}, the leading order operators for FSR are dimension 6 operators of the form $(\chi^\dagger\bar{\sigma}^\mu\chi)(e^\dagger\bar{\sigma}_\mu e)$.  Again, these operators are chirality-suppressed in the s-wave approximation, thus the related FSR cross-section is
\begin{equation}
\langle\sigma_{\rm FSR}v\rangle\approx c_{\rm FSR}\alpha\frac{m_e^2}{M_{\rm int}^4}\ln\left[\frac{4m_\chi^2}{m_e^2}\right]
\end{equation}
where $c_{\rm FSR}$ includes the operator coupling constant and the $\pi$ factors from the phase space integration.

Combining operators of table \ref{tab:opsMajDMann} with operators of tables \ref{tab:opsElectron} and \ref{tab:opsPhoton}, the leading operators for direct photon production are dimension 8 and are not chirality-suppressed.  The operator made out of $(\chi^\dagger\bar{\sigma}^\mu\chi)$, $(e^\dagger\bar{\sigma}^\nu e)$ and $F_{\mu\nu}$ is an example.  The related direct photon production cross-section is
\begin{equation}
\langle\sigma_{\rm eff}v\rangle\approx c_{\rm eff}\alpha\frac{(2m_\chi)^6}{M_{\rm int}^8}
\end{equation}
where $c_{\rm eff}$ includes the operator coupling constant and the $\pi$ factors from the phase space integration.

Comparing FSR and direct photon production cross-sections leads to the equivalent of equation (\ref{eqn:FSRvsOPS}) with the same overall conclusions as scalar boson DM annihilation.

\subsection*{Dirac DM annihilation to $e^++e^-+\gamma$}

The relevant lowest-dimension Dirac operators are given in table \ref{tab:opsDiracDMann}.
\begin{table}[ht]
\begin{center}
\includegraphics[scale=0.7]{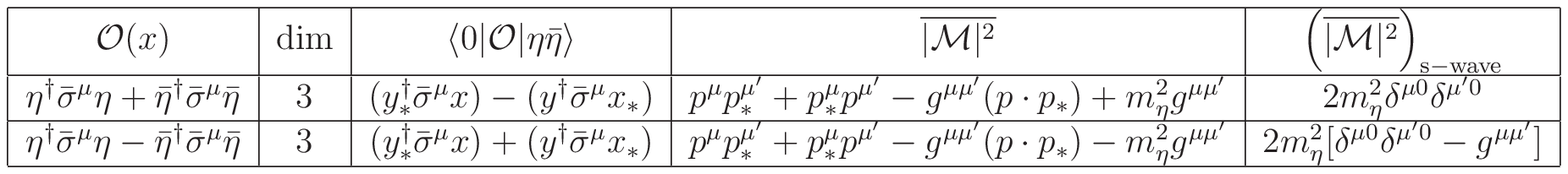}
\caption{Relevant Majorana DM operators for DM annihilation.}
\label{tab:opsDiracDMann}
\end{center}
\end{table}
Combining operators of table \ref{tab:opsDiracDMann} with operators of table \ref{tab:opsElectron}, the leading order operators for FSR are dimension 6 operators of the form $[(\eta^\dagger\bar{\sigma}^\mu\eta)-(\bar{\eta}^\dagger\bar{\sigma}^\mu\bar{\eta})](e^\dagger\bar{\sigma}_\mu e)$.  However, these operators are not chirality-suppressed in the s-wave approximation, thus the related FSR cross-section is
\begin{equation}
\langle\sigma_{\rm FSR}v\rangle\approx c_{\rm FSR}\alpha\frac{4m_\eta^2}{M_{\rm int}^4}\ln\left[\frac{4m_\eta^2}{m_e^2}\right]
\end{equation}
where $c_{\rm FSR}$ includes the operator coupling constant and the $\pi$ factors from the phase space integration.

Combining operators of table \ref{tab:opsDiracDMann} with operators of tables \ref{tab:opsElectron} and \ref{tab:opsPhoton}, the leading operators for direct photon production are dimension 8 and are not chirality-suppressed, as in the Majorana DM annihilation case.  The operator made out of $[(\eta^\dagger\bar{\sigma}^\mu\eta)+(\bar{\eta}^\dagger\bar{\sigma}^\mu\bar{\eta})]$, $(e^\dagger\bar{\sigma}^\nu e)$ and $F_{\mu\nu}$ is an example.  The related direct photon production cross-section is
\begin{equation}
\langle\sigma_{\rm eff}v\rangle\approx c_{\rm eff}\alpha\frac{(2m_\eta)^6}{M_{\rm int}^8}
\end{equation}
where $c_{\rm eff}$ includes the operator coupling constant and the $\pi$ factors from the phase space integration.

Comparing FSR and direct photon production cross-sections leads to
\begin{equation}
\langle\sigma_{\rm FSR}v\rangle\approx\langle\sigma_{\rm eff}v\rangle\frac{c_{\rm FSR}}{c_{\rm eff}}\frac{M_{\rm int}^4}{(2m_\eta)^4}\ln\left[\frac{4m_\eta^2}{m_e^2}\right]\gg\langle\sigma_{\rm eff}v\rangle
\end{equation}
and, for order one coefficients, FSR always dominates over direct photon production from higher-order operators.

\subsection*{Gauge boson DM annihilation to $e^++e^-+\gamma$}

The relevant lowest-dimension gauge boson operators are given in table \ref{tab:opsGaugeDMann}.
\begin{table}[ht]
\begin{center}
\includegraphics[scale=0.65]{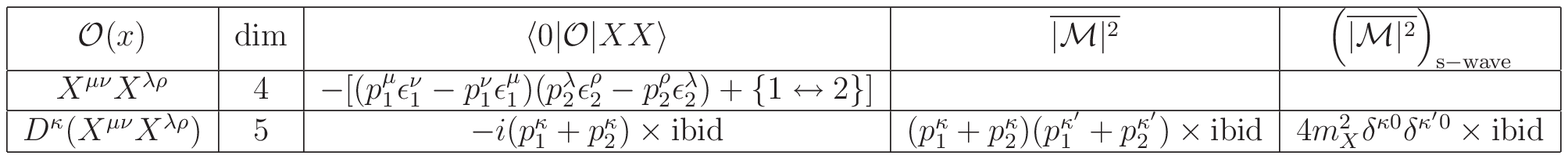}
\caption{Relevant gauge boson DM operators for DM annihilation.}
\label{tab:opsGaugeDMann}
\end{center}
\end{table}
Combining operators of table \ref{tab:opsGaugeDMann} with operators of table \ref{tab:opsElectron}, the leading order operators for FSR are dimension 8 operators of the form $X^{\mu\lambda}X_{\nu\lambda}i(D^\nu e^\dagger\bar{\sigma}_\mu e-e^\dagger\bar{\sigma}_\mu D^\nu e)$.  However, these operators are not chirality-suppressed in the s-wave approximation, thus the related FSR cross-section is
\begin{equation}
\langle\sigma_{\rm FSR}v\rangle\approx c_{\rm FSR}\alpha\frac{(2m_X)^6}{M_{\rm int}^8}\ln\left[\frac{4m_X^2}{m_e^2}\right]
\end{equation}
where $c_{\rm FSR}$ includes the operator coupling constant and the $\pi$ factors from the phase space integration.

Combining operators of table \ref{tab:opsGaugeDMann} with operators of tables \ref{tab:opsElectron} and \ref{tab:opsPhoton}, the leading operators for direct photon production are dimension 8 and are not chirality-suppressed.  The operator made out of $X^{\mu\lambda}X_{\nu\lambda}$ and $i(D^\nu e^\dagger\bar{\sigma}_\mu e-e^\dagger\bar{\sigma}_\mu D^\nu e)$ is an example.  The related direct photon production cross-section is
\begin{equation}
\langle\sigma_{\rm eff}v\rangle\approx c_{\rm eff}\alpha\frac{(2m_X)^6}{M_{\rm int}^8}
\end{equation}
where $c_{\rm eff}$ includes the operator coupling constant and the $\pi$ factors from the phase space integration.

Comparing FSR and direct photon production cross-sections leads to
\begin{equation}
\langle\sigma_{\rm FSR}v\rangle\approx\langle\sigma_{\rm eff}v\rangle\frac{c_{\rm FSR}}{c_{\rm eff}}\ln\left[\frac{4m_X^2}{m_e^2}\right]\gtrsim\langle\sigma_{\rm eff}v\rangle
\end{equation}
and, for order one coefficients, FSR always dominates over direct photon production from higher-order operators, although only slightly.

\subsection*{Scalar boson DM decay to $e^++e^-+\gamma$}

This case follows closely the scalar boson DM annihilation case.  The relevant lowest-dimension scalar boson operators are given in table \ref{tab:opsScalarDMdecay}.
\begin{table}[ht]
\begin{center}
\includegraphics[scale=1]{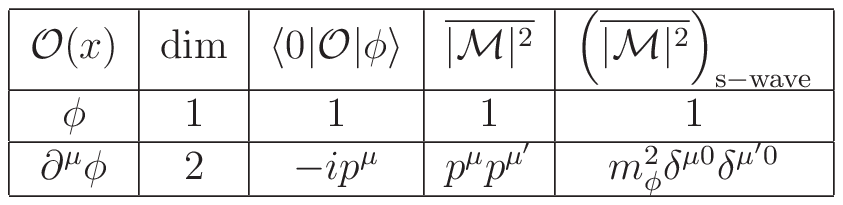}
\caption{Relevant scalar boson DM operators for DM decay.}
\label{tab:opsScalarDMdecay}
\end{center}
\end{table}
Combining operators of table \ref{tab:opsScalarDMdecay} with operators of table \ref{tab:opsElectron}, the leading order operators for FSR are dimension 5 operators of the form $m_e\phi(\bar{e}e+e^\dagger\bar{e}^\dagger)$.  These operators are chirality-suppressed and thus the related FSR decay rate is
\begin{equation}
\Gamma_{\rm FSR}\approx c_{\rm FSR}\alpha\frac{m_e^2m_\phi}{M_{\rm int}^2}\ln\left[\frac{m_\phi^2}{m_e^2}\right]
\end{equation}
where $c_{\rm FSR}$ includes the operator coupling constant and the $\pi$ factors from the phase space integration.

Combining operators of table \ref{tab:opsScalarDMdecay} with operators of tables \ref{tab:opsElectron} and \ref{tab:opsPhoton}, the leading operators for direct photon production are dimension 7 and are not chirality-suppressed.  The operator made out of $\partial^\mu\phi$, $(e^\dagger\bar{\sigma}^\nu e)$ and $F_{\mu\nu}$ is an example.  The related direct photon production decay rate is
\begin{equation}
\Gamma_{\rm eff}\approx c_{\rm eff}\alpha\frac{m_\phi^7}{M_{\rm int}^6}
\end{equation}
where $c_{\rm eff}$ includes the operator coupling constant and the $\pi$ factors from the phase space integration.

Comparing FSR and direct photon production decay rates leads to the equivalent of equation (\ref{eqn:FSRvsOPS}) for decay rates,
\begin{equation}
\Gamma_{\rm FSR}\approx\Gamma_{\rm eff}\frac{c_{\rm FSR}}{c_{\rm eff}}\frac{m_e^2M_{\rm int}^4}{(2m_\phi)^6}\ln\left[\frac{4m_\phi^2}{m_e^2}\right],
\end{equation}
with the same overall conclusions as scalar boson DM annihilation.

\subsection*{Abelian gauge boson DM decay to $e^++e^-+\gamma$}

The relevant lowest-dimension gauge boson operators are given in table \ref{tab:opsGaugeDMdecay}.
\begin{table}[ht]
\begin{center}
\includegraphics[scale=1]{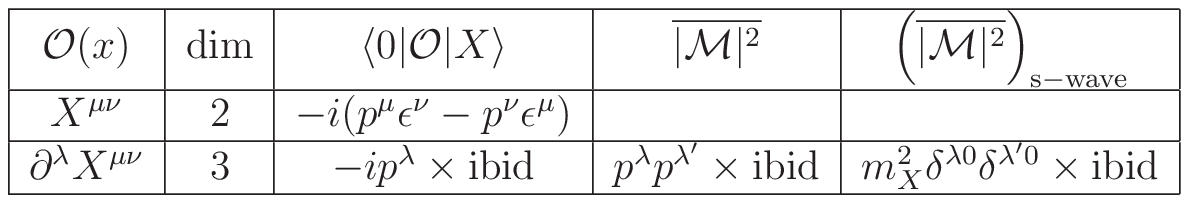}
\caption{Relevant gauge boson DM operators for DM decay.}
\label{tab:opsGaugeDMdecay}
\end{center}
\end{table}
Combining operators of table \ref{tab:opsGaugeDMdecay} with operators of table \ref{tab:opsElectron}, the leading order operators for FSR are dimension 6 operators of the form $\partial^\mu X_{\mu\nu}(e^\dagger\bar{\sigma}^\nu e)$.  However, these operators are not chirality-suppressed and thus the related FSR decay rate is
\begin{equation}
\Gamma_{\rm FSR}\approx c_{\rm FSR}\alpha\frac{m_X^5}{M_{\rm int}^4}\ln\left[\frac{m_X^2}{m_e^2}\right]
\end{equation}
where $c_{\rm FSR}$ includes the operator coupling constant and the $\pi$ factors from the phase space integration.

Combining operators of table \ref{tab:opsGaugeDMdecay} with operators of tables \ref{tab:opsElectron} and \ref{tab:opsPhoton}, the leading operators for direct photon production are dimension 6 and are not chirality-suppressed.  The operator made out of $X^{\mu\nu}$ and $i(D_ \nu e^\dagger\bar{\sigma}_\mu e-e^\dagger\bar{\sigma}_\mu D_ \nu e)$ is an example.  The related direct photon production decay rate is
\begin{equation}
\Gamma_{\rm eff}\approx c_{\rm eff}\alpha\frac{m_X^5}{M_{\rm int}^4}
\end{equation}
where $c_{\rm eff}$ includes the operator coupling constant and the $\pi$ factors from the phase space integration.

Comparing FSR and direct photon production cross-sections leads to
\begin{equation}
\Gamma_{\rm FSR}\approx\Gamma_{\rm eff}\frac{c_{\rm FSR}}{c_{\rm eff}}\ln\left[\frac{m_X^2}{m_e^2}\right]\gtrsim\Gamma_{\rm eff}
\end{equation}
and, for order one coefficients, FSR always dominates over direct photon production from higher-order operators, although only slightly.

\subsection*{DOS for direct photon production from higher-order operators: An example for Majorana DM annihilation}

When direct photon production from higher-order operators dominates over FSR, the photon DOS can be obtained from the most general effective Lagrangian.  Here we study the case of Majorana DM annihilation to electron-positron pair, $\chi+\chi\rightarrow e^++e^-+\gamma$.  The other cases where direct photon production from higher-order operators dominates over FSR are basically equivalent.

The lowest-dimension operators relevant to $\chi+\chi\rightarrow e^++e^-+\gamma$ are effective dimension 8 operators due to the chirality rule, thus only effective mass dimension 8 operators will be considered.  The highest possible effective mass dimension for the DM ($\chi$) operators is therefore 4.  The photon can either appear in a covariant derivative or in the field strength tensor.

With the help of tables \ref{tab:opsElectron}, \ref{tab:opsPhoton} and \ref{tab:opsMajDMann}, the minimal set (in the sense that any operator relevant for $\chi+\chi\rightarrow e^++e^-+\gamma$ can be rewritten as a linear combination of the operators in the minimal set) of operators of the mass dimension 8 effective Lagrangian for the process $\chi+\chi\rightarrow e^++e^-+\gamma$ can be found.

From the electron operators without covariant derivatives, only the one with mass dimension 3 is relevant, leading to $(\chi^\dagger\bar{\sigma}^\mu\chi)(e^\dagger\bar{\sigma}^\nu e)F^{\lambda\rho}$.  From the electron operators with one covariant derivative, only $i(\chi^\dagger\bar{\sigma}^\mu\chi)\partial^\rho(D^\lambda e^\dagger\bar{\sigma}^\nu e-e^\dagger\bar{\sigma}^\nu D^\lambda e)$ survives (there are two different ways of building this operator) since all other operators are either zero to lowest order in $m_e$ from the equations of motion (the $\sigma$ matrix and the covariant derivative are forced to be contracted together) or vanish in the s-wave.  Finally there are only two operators that can be built from the electron operators with two covariant derivatives, which are $(\chi^\dagger\bar{\sigma}^\mu\chi)(D^\rho D^\lambda e^\dagger\bar{\sigma}^\nu e+e^\dagger\bar{\sigma}^\nu D^\rho D^\lambda e)$ and $i(\chi^\dagger\bar{\sigma}^\mu\chi)(D^\rho D^\lambda e^\dagger\bar{\sigma}^\nu e-e^\dagger\bar{\sigma}^\nu D^\rho D^\lambda e)$.

All operators have four Lorentz indices and must therefore be contracted with $g_{\mu\nu}g_{\lambda\rho}$, $g_{\mu\lambda}g_{\nu\rho}$, $g_{\mu\rho}g_{\nu\lambda}$ and $\epsilon_{\mu\nu\lambda\rho}$.  Using the equations of motion, this fact leads to an even smaller minimal set (to lowest order in $m_e$), since $g_{\mu\nu}1=\sigma_\mu\bar{\sigma}_\nu+2i\sigma_{\mu\nu}$ and $[D_\mu,D_\nu]=-i\sqrt{4\pi\alpha}F_{\mu\nu}$.  Indeed one has
\begin{eqnarray*}
g_{\lambda\rho}(e^\dagger\bar{\sigma}^\nu D^\rho D^\lambda e) &=& e^\dagger\bar{\sigma}^\nu(\sigma_\rho\bar{\sigma}_\lambda D^\rho D^\lambda+g\sigma_{\rho\lambda}F^{\rho\lambda})e\sim F^{\lambda\rho}{\rm -term}+\mathcal{O}(m_e)\\
g_{\nu\rho}(e^\dagger\bar{\sigma}^\nu D^\rho D^\lambda e) &=& g_{\nu\rho}e^\dagger\bar{\sigma}^\nu(D^\lambda D^\rho+[D^\rho,D^\lambda])e\sim F^{\lambda\rho}{\rm -term}+\mathcal{O}(m_e)\\
g_{\nu\lambda}(e^\dagger\bar{\sigma}^\nu D^\rho D^\lambda e) &\sim& \mathcal{O}(m_e)\\
\epsilon_{\mu\nu\lambda\rho}(e^\dagger\bar{\sigma}^\nu D^\rho D^\lambda e) &=& \frac{1}{2}\epsilon_{\mu\nu\lambda\rho}(e^\dagger\bar{\sigma}^\nu[D^\rho,D^\lambda]e)\sim F^{\lambda\rho}{\rm -term}
\end{eqnarray*}
thus eliminating all operators with two covariant derivatives.  Moreover, the operator with one covariant derivative can be rewritten in terms of two covariant derivatives as $i(\chi^\dagger\bar{\sigma}^\mu\chi)(D^\rho e^\dagger\bar{\sigma}^\nu D^\lambda e-D^\lambda e^\dagger\bar{\sigma}^\nu D^\rho e)$, leading to
\begin{eqnarray*}
g_{\lambda\rho}(D^\rho e^\dagger\bar{\sigma}^\nu D^\lambda e-D^\lambda e^\dagger\bar{\sigma}^\nu D^\rho e) &=& 0\\
g_{\nu\rho}(D^\rho e^\dagger\bar{\sigma}^\nu D^\lambda e-D^\lambda e^\dagger\bar{\sigma}^\nu D^\rho e) &\sim& \mathcal{O}(m_e)\\
g_{\nu\lambda}(D^\rho e^\dagger\bar{\sigma}^\nu D^\lambda e-D^\lambda e^\dagger\bar{\sigma}^\nu D^\rho e) &\sim& \mathcal{O}(m_e)\\
\epsilon_{\mu\nu\lambda\rho}(D^\rho e^\dagger\bar{\sigma}^\nu D^\lambda e-D^\lambda e^\dagger\bar{\sigma}^\nu D^\rho e) &=& 2\,\epsilon_{\mu\nu\lambda\rho}(D^\rho e^\dagger\bar{\sigma}^\nu D^\lambda e)\\
 &=& 2\,\epsilon_{\mu\nu\lambda\rho}[\partial^\rho(e^\dagger\bar{\sigma}^\nu D^\lambda e)-(e^\dagger\bar{\sigma}^\nu D^\rho D^\lambda e)]\\
 &\sim& \epsilon_{\mu\nu\lambda\rho}\partial^\rho(e^\dagger\bar{\sigma}^\nu D^\lambda e)+F^{\lambda\rho}{\rm -term}
\end{eqnarray*}
therefore eliminating three of the four possible operators.  Finally, for the operators with $F^{\mu\nu}$, only two operators survive since the field strength tensor is antisymmetric.

Then the minimal set consists of $(\chi^\dagger\bar{\sigma}^\mu\chi)(e^\dagger\bar{\sigma}^\nu e)F_{\mu\nu}$, $(\chi^\dagger\bar{\sigma}^\mu\chi)(e^\dagger\bar{\sigma}^\nu e)\tilde{F}_{\mu\nu}$ and $i\epsilon_{\mu\nu\lambda\rho}\partial^\rho(\chi^\dagger\bar{\sigma}^\mu\chi)(e^\dagger\bar{\sigma}^\nu D^\lambda e)$.  In the s-wave approximation, the last term also vanishes, thus leading to only two operators relevant for the mass dimension 8 effective Lagrangian of $\chi+\chi\rightarrow e^++e_-+\gamma$ (at lowest order in $m_e$),
\begin{equation}
\mathcal{L}_{\rm eff}=\frac{\sqrt{4\pi\alpha}}{M_{\rm int}^4}(\chi^\dagger\bar{\sigma}^\mu\chi)(e^\dagger\bar{\sigma}^\nu e)[a_LF_{\mu\nu}+b_L\tilde{F}_{\mu\nu}]+\{L\rightarrow R,e\rightarrow\bar{e}\}
\end{equation}
where the coupling constants are assumed to be order one numbers.  Since the operators do not interfere (they couple the electrons to different photon states), the probability is simply given by
\begin{eqnarray*}
\overline{|\mathcal{M}|^2} &=& -32\pi\alpha(a_L^2+b_L^2+a_R^2+b_R^2)m_\chi^2M_{\rm int}^{-8}\\
    && \hspace{1cm}\times[(p_\gamma\cdot p_+)(p_\gamma\cdot p_-)-E_\gamma E_+(p_\gamma\cdot p_-)-E_\gamma E_-(p_\gamma\cdot p_+)]\\
 &=& 64\pi\alpha(a_L^2+b_L^2+a_R^2+b_R^2)m_\chi^3(m_\chi-E_\gamma)M_{\rm int}^{-8}\\
 && \hspace{1cm}\times(2m_\chi^2-2m_\chi E_\gamma-4m_\chi E_++E_\gamma^2+2E_\gamma E_++2E_+^2)
\end{eqnarray*}
where $\vec{p}_-=-\vec{p}_\gamma-\vec{p}_+$, $E_-=\sqrt{E_\gamma^2+E_+^2+2zE_\gamma E_+}=2m_\chi-E_\gamma-E_+$, $z=\frac{2m_\chi^2-2m_\chi E_\gamma-2m_\chi E_++E_\gamma E_+}{E_\gamma E_+}$ and $m_\chi-E_\gamma<E_+<m_\chi$.  Here $z=\cos\theta$ and $\theta$ is the angle between the positron and the photon.  In the vanishing electron mass limit the annihilation cross-section is thus
\begin{equation}
\frac{d\langle\sigma v\rangle}{dE_\gamma}=\frac{1}{4m_\chi^2}\int_{m_\chi-E_\gamma}^{E_\gamma}\frac{dE_+}{32\pi^3}\overline{|\mathcal{M}|^2}=\frac{\alpha(a_L^2+b_L^2+a_R^2+b_R^2)m_\chi}{3\pi^2M_{\rm int}^8}(m_\chi-E_\gamma)E_\gamma^3
\end{equation}
with $0<E_\gamma<m_\chi$, which gives the DOS mentioned in the text.  Notice that the annihilation cross-section vanishes like the cube of the photon energy, $E_\gamma^3$, as $E_\gamma\rightarrow0$.


\end{document}